\documentclass[fleqn,usenatbib]{mnras}

\usepackage[T1]{fontenc}

\DeclareRobustCommand{\VAN}[3]{#2}
\let\VANthebibliography\thebibliography
\def\thebibliography{\DeclareRobustCommand{\VAN}[3]{##3}\VANthebibliography}

\usepackage{graphicx}	
\usepackage{amsmath}	
\usepackage{amssymb}	
\usepackage{array}
\usepackage{tabularx}
\usepackage{makecell}
\usepackage{xcolor} 

\usepackage{newtxtext,newtxmath}

\newcommand\kev{{\rm~keV}}

\newcommand\kms{\ifmmode {\rm~km\ s}^{-1} \else ~km s$^{-1}$\fi}
\newcommand\Hunit{\ifmmode {\rm~km\ s}^{-1}\ {\rm Mpc}^{-1}
        \else ~km s$^{-1}$ Mpc$^{-1}$\fi}
\newcommand\ctssec{\ifmmode {\rm~count\ s}^{-1} \else ~count s$^{-1}$\fi}
\newcommand\ergsec{\ifmmode {\rm~erg\ s}^{-1} \else
        ~erg s$^{-1}$\fi}
\newcommand\funit{\ifmmode {\rm~erg\ s}^{-1}\;{\rm cm}^{-2} \else
        ~ergs s$^{-1}$ cm$^{-2}$\fi}
\newcommand\phflux{\ifmmode {\rm~photon\ s}^{-1}\;{\rm cm}^{-2}
        \else   ~photon s$^{-1}$ cm$^{-2}$\fi}
\newcommand\efluxA{\ifmmode {\rm~erg\ s}^{-1}\;{\rm cm}^{-2}\;{\rm
        \AA}^{-1} \else ~erg s$^{-1}$ cm$^{-2}$ \AA$^{-1}$\fi}
\newcommand\efluxHz{\ifmmode {\rm~erg\ s}^{-1}\;{\rm cm}^{-2}\;{\rm
        Hz}^{-1} \else ~erg s$^{-1}$ cm$^{-2}$ Hz$^{-1}$\fi}
\newcommand\cc{\ifmmode {\rm~cm}^{-3} \else cm$^{-3}$\fi}
\newcommand\FWHM{\ifmmode {\rm~FWHM} \else ${\rm~FWHM}$\fi}
\newcommand\Zsun{\ifmmode Z_{\odot} \else $M_{\odot}$\fi}
\newcommand\Lsun{\ifmmode L_{\odot} \else $L_{\odot}$\fi}
\newcommand\ltsim{\raisebox{-.5ex}{$\;\stackrel{<}{\sim}\;$}}
\newcommand\gtsim{\raisebox{-.5ex}{$\;\stackrel{>}{\sim}\;$}}
\newcommand\hbeta{\ifmmode {\rm H}\beta \else H$\beta$\fi}
\newcommand\Kalpha{\ifmmode {\rm K}\alpha \else K$\alpha$\fi}
\newcommand\nh{\ifmmode N_{\rm H} \else N$_{\rm H}$\fi}

\newcommand{\lum}{erg\,s$^{-1}$}

\newcommand{\Msun}{\ensuremath{\rm M_{\odot}}}
\newcommand{\sersic}{S\'{e}rsic}

\title[{}HST study of  UGC 12591]{{\it Hubble Space
Telescope} Captures UGC~12591: Bulge/Disc  Properties, Star Formation  and  `Missing Baryons' Census in  a Very Massive and Fast Spinning Hybrid Galaxy}

\author[Ray et al.]{
Shankar Ray,$^{1,2}$\thanks{E-mail: rays71349@gmail.com}
Joydeep Bagchi,$^{2}$\thanks{E-mail: joydeep.bagchi@christuniversity.in}
Suraj Dhiwar,$^{1,3,4}$
Pandge, M. B.$^{1}$\thanks{E-mail: mbpandge@gmail.com}
Mohammad Mirakhor,$^{5}$ 
\newauthor
 Stephen A. Walker,$^{5}$
Dipanjan  Mukherjee$^{3}$
\\
$^{1}$Dayanand Science College,
 Latur 413531, India\\
$^{2}$ Department of Physics \& Electronics, CHRIST (Deemed to be University), Hosur Road, Bengaluru 560029, India\\
$^{3}$ Inter-University Centre for Astronomy and Astrophysics,
 Pune 411007, India\\
$^{4}$ Savitribai Phule Pune University, Pune 411007, India\\
$^{5}$ The University of Alabama in Huntsville,  OPB430
Huntsville AL, 35899, U.S.A.\\
}

\date{Accepted XXX. Received YYY; in original form ZZZ}

\pubyear{2021}

\begin{document}
\label{firstpage}
\pagerange{\pageref{firstpage}--\pageref{lastpage}}
\maketitle

\begin{abstract}   

 We present Hubble  Space Telescope (HST) observations of the nearby, massive, highly rotating hybrid galaxy UGC~12591, along with observations in UV to FIR bands. HST data in V, I, and H bands is used to disentangle the structural components. Surface photometry shows a dominance of the bulge over the disc with H-band B/D ratio of $69\%$. The spectral energy distribution (SED) fitting reveals an extremely  low global star formation rate (SFR) of  $\rm\sim0.1-0.2  M_\odot yr^{-1}$, exceptionally  low for the galaxy's huge stellar mass of $\rm 1.6\times10^{11}M_\odot$, implying  a  strong quenching of its SFR with star formation efficiency of $3-5\%$.  For at least the past  $\rm 10^{8}$ years, the galaxy has remained in a quiescent state as a sterile,  `red and dead' galaxy. UGC~12591 hosts a supermassive black hole (SMBH) of $\rm 6.18\times 10^{8} M_\odot$ which is possibly quiescent at present, i.e. neither we see large  ($\rm>1 kpc$) radio jets nor is the  SMBH contributing significantly to the  mid-IR SED,  ruling out strong  radiative feedback of AGN.
We obtained a  detailed census of all observable baryons with a total mass of $\rm 6.46\times10^{11} M_\odot$ within the virial radius, amounting to a  baryonic deficiency of  $\sim$$85\%$ relative to the cosmological mean. Only a small fraction of these baryons resides in a warm/hot circum-galactic X-ray halo, while the majority are still unobservable. We discussed  various astrophysical 
scenarios for explaining its unusual properties. Our work is a major step forward in understanding  the assembly history of such extremely massive,  isolated galaxies.


\end{abstract}

\begin{keywords}
galaxies: ISM $-$ Galaxy: formation $-$ galaxies: spiral $-$ Galaxy: fundamental parameters $-$ galaxies: individual: UGC12591
\end{keywords}



\section{Introduction}
\label{sec:intro}

Explaining galaxy formation in the Universe across cosmic time,  within the  current, widely accepted Cosmological paradigm  known as
the concordance $\rm \Lambda$CDM cosmology,  is one of the  most important goals of  astronomy research.
It is believed that a number  of physical processes are responsible for the observed properties of galaxies, and identifying the correct ones among them is one of the principal goals of  current  galaxy formation theory. 
Star formation 
rate (SFR) reached  it's peak at `cosmic high noon' near $\it z = 2$ when galaxies rapidly built up their
stellar masses by converting  cosmic baryons to stars, and subsequently, over billions of years the
SFR went down drastically, and by the present era most of the massive
galaxies have stopped growing, entering a 
quiescent state \citep[][]{2010ApJ...717..379B,2019MNRAS.488.3143B,2013MNRAS.428.3121M}. For reasons still unclear, even now  some lower mass galaxies are still actively forming stars and massive local galaxies with halo mass $\gtsim 10^{12} \Msun$ have low specific star formation rate compared to less massive spirals (see \citealt{2018ARA&A..56..435W} for a review). 

The formation era of  progenitors of these massive galaxies is debatable,  but it  could stretch  as far back in time as $\rm {\it z} > 6$,  when they  gradually grew in mass as well as formed new stars from  cosmic baryons. Near redshift $\rm {\it z} = 2$, when the Universe was about three billion years old, half of the most massive galaxies were in place as compact proto-galaxies  and they had already exhausted  their fuel for star formation. Of the rest, about $25\%$  galaxies assembled before the peak of the cosmic star-formation,  while  $25\%$ galaxies formed later \citep[][]{2014ARA&A..52..415M}. At higher redshifts, it is observed  that  most of them  were harboring  intense nuclear starbursts and possibly  they ultimately grew into the most  massive local  galaxies seen today,  through mergers with minor companions and gas accretion from cosmic-web filaments. Unveiling how this remarkable transformation happened  and what physical factors were behind this change has been an enduring quest,  which  brings forth   some of   most  hotly debated,  key  questions in galaxy formation (for reviews  see \citet[][]{2020NatRP...2...42V} and \citet[][]{2015ARA&A..53...51S}).

One of the prime  mechanism of suppression and  even halting of star formation  in  galaxies is  believed to have operated via a  process called AGN (Active Galactic Nuclei) feedback. When  supermassive black holes (SMBH) growing at the centres of galaxies accrete matter, they turn into powerful engines \citep[][]{1969Natur.223..690L}  that can potentially heat and even expel the  gas of their host galaxies, thereby  halting star formation  \citep[][]{2008MNRAS.391..481S,2005Natur.433..604D,2017MNRAS.472..949B}. 
It is now widely recognized that 
in our local Universe such  SMBHs of mass range $\rm \sim 10^{6} - 10^{10} \Msun$ lurk in the nuclei of almost all massive galaxies (see  \citet[]{2013ARA&A..51..511K} for a review). 
AGN feedback is likely to be more effective in   massive galaxies
because a remarkable tight coupling exists  between the mass of the black hole
and the galaxy's  bulge and  the  total  stellar mass/luminosity, which suggests that; (a) the mass of the central SMBH will be higher in massive bulge dominated galaxies and so will its energetic output and  impact on the surroundings (b) 
the star formation  rate  and  final mass of the galaxy has  somehow been regulated 
symbiotically  by the  black hole over cosmic time, e.g., \citep[][]{1998AJ....115.2285M,2000ApJ...539L..13G,2004ApJ...604L..89H,2003ApJ...589L..21M,2009ApJ...698..198G,2019ApJ...876..155S}. In spite of the popularity
of AGN feedback models, there is no good understanding of exactly how the  AGN is fuelled 
and how the radio jets or hot outflows from the SMBHs deposit   energy to their surroundings, the energy that is  transferred to the different gas phases, and  what are the physical conditions when significant positive or negative feedback  on star formation rate  is produced (see review by \citet[][]{2012ARA&A..50..455F}).
Despite numerous efforts to observe SMBHs in the process of quenching star formation, conclusive evidence for such a process has remained  elusive, particularly in the nearby Universe. The impact of AGN feedback via radio jet mode on the hot intracluster medium (ICM) is  observed directly through X-ray observations of the cavities and bubbles around the central galaxies of  cool core clusters.

Some observations have recently  confirmed AGN feedback scenario locally  for galaxies, e.g.  \citep[][]{2018Natur.553..307M}. 
Recent Carbon Monoxide and radio observations of a  rare, super   massive, jetted  spiral galaxy 2MASX J23453268-0449256 (J2345-0449 hereafter)  \citep[][]{2014ApJ...788..174B,2021A&A...654A...8N} and  
relativistic hydrodynamic simulations of radio jets propagating in a dense ISM
\citep[][]{2016MNRAS.461..967M,2021MNRAS.508.4738M}  provide strong
support for AGN-jet feedback scenario.  Interstellar turbulence may also play a major
role in quenching star formation in AGN hosted galaxies, as shown by observations
and recent hydrodynamic simulations \citep[][]{2010A&A...521A..65N,2016ApJ...826...29L,2012ApJ...761..156F,2021MNRAS.508.4738M,2016MNRAS.461..967M}.

The Circum Galactic Medium (CGM), around galaxies plays an important role in deciding, at all cosmological epochs, how galaxies acquire, eject, and recycle their gas, which in turn  are the most important physical factors in  determining  how the galaxies evolve,  how much star formation will take place,  or  even  how much  of it will  be  quenched or is recycled back.  Apart from the unseen dark-matter, the CGM is an extremely dynamic environment having a complex mix of  baryons existing  in various  phases of  temperature, density and  metal content,  which  presents  an  extremely important but perhaps the  least understood aspect of galaxy formation (see \cite[][]{2017ARA&A..55..389T} for a review). Among the many interesting and challenging problems that the  CGM presents,  one of the most perplexing  has been the so called  `missing baryons'  problem. The term `missing baryons' refers to the fact that even in multi-wavelength observations of  galaxies, the total  detectable baryonic mass falls far short (by 50 to 80 $\%$) of the  cosmic baryon fraction  of $\rm \approx 0.15 - 0.17$ as expected in $\Lambda$CDM cosmology, based on Big Bang nucleo-synthesis and Cosmic Microwave Background constraints (see
\citet[][]{2007ARA&A..45..221B} for a review and references therein). A number of  theories and simulations have tried to explain  why only a small fraction of theses halo baryons condenses into stars and why by $\rm {\it z} = 0$ the star formation efficiency $\rm f_{\star}$ reaches a peak  of  about $\rm  20 - 30 \%$ for  typical $\rm L_{\star}$ galaxies. For very massive galaxies (halo mass $\rm \gtsim M_{h} \sim 10^{12} M_{\sun}$), beyond the peak,  $\rm f_{\star}$ declines rapidly to  $\rm f_{\star} \sim 2 - 5 \%$ at $\rm  M_{h} \sim 10^{13} M_{\sun}$ \citep[][]{2015MNRAS.446..521S,2010ApJ...717..379B,2019MNRAS.488.3143B,2013MNRAS.428.3121M}. It is believed  that AGN feedback is one of the main factors needed to bring this  decline in SFR of most massive galaxies. A similar decline in SFR is seen on the lower mass end of the peak ($\rm \ltsim M_{h} \sim 10^{12} M_{\sun}$), attributed to supernova feedback. 

Deep study 
 of gas phases  at large distances in CGM may  provide  powerful constraints on the  feedback processes.
Leading galaxy formation theories \citep[][]{1978MNRAS.183..341W,1991ApJ...379...52W,2006ApJ...639..590F,2006ApJ...644L...1S} demand a warm-hot X-ray corona containing a large fraction  (at least $50 \%$) of cosmic baryons in such massive galaxies. Observing such baryon filled halos have been elusive so far. To validate galaxy formation models and make a census of baryons we need sensitive X-ray or UV observations of galactic halos.  
It has been proved very challenging to detect the hot gaseous corona  around  even very luminous spirals and ellipticals with our present telescopes due to sensitivity limitations. With XMM-Newton such coronas have  been robustly detected in only a handful of
galaxies so far, e.g., \citep[][]{2012ApJ...755..107D,2021MNRAS.500.2503M,
2015MNRAS.449.3527W,2011ApJ...737...22A,2016MNRAS.455..227A,2018ApJ...862....3B,2013ApJ...772...97B} and independent constraints on their temperature, density, and
metallicity  from soft X-ray spectroscopy is still lacking. These galaxies provide compelling evidence for the existence of hot, low-metallicity atmospheres of gas that
could originate with accretion from the IGM and subsequent  heating to the virial temperature of the halo via accretion shocks. However,  the fraction of baryons residing in the hot phase, and its dependence on stellar and or halo mass, as well as
how they are affected by galactic feedback and star formation processes are not  well
determined and  poorly understood so far (see \citet[][]{2017ARA&A..55..389T}).

\begin{figure*}
\begin{center}
\includegraphics[scale=.38]{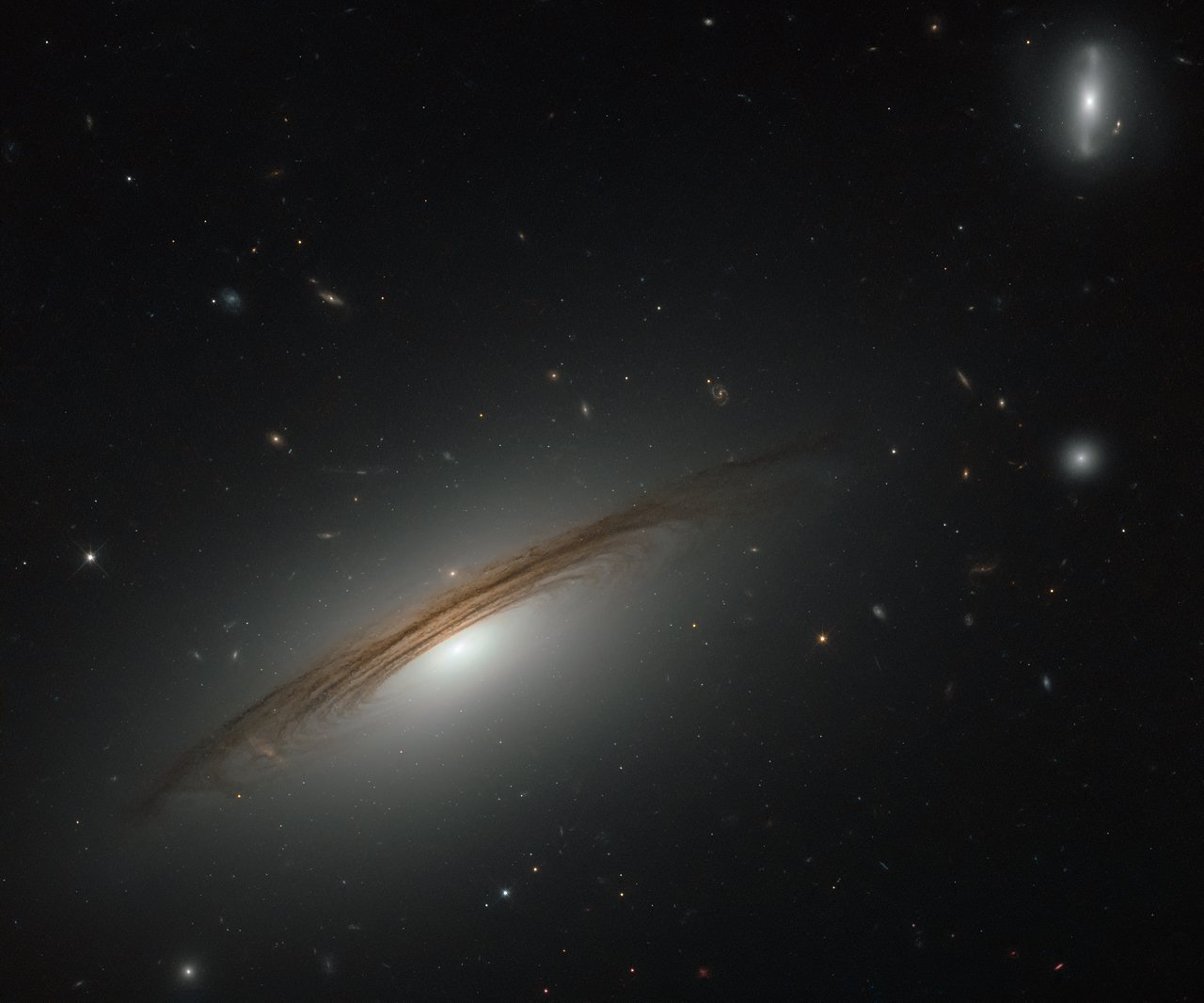}
\end{center}
\caption{A colour image of the  UGC~12591 is shown as imaged
by {\it Hubble Space Telescope}.  A massive stellar bulge and prominent, stratified  dark dust lanes can clearly be seen crossing the equatorial plane and a  bright, compact central nucleus is also noticeable. The barred spindle shaped satellite galaxy in the upper right corner is
WISEA J232529.77+283021.3 at redshift 0.0233, the same as UGC~12591. Image credit: (ESA/Hubble and NASA)}
\label{fig:HST}
\end{figure*}

\vskip 0.5cm
Our target
UGC~12591 (23h 25m 21.75s, +28d 29m 42.76s J2000.0)
at a redshift of ${\it z}\!=\!0.023$ is classified as a S0/Sa type isolated galaxy located at the western-most fringes of Perseus-Pisces supercluster. It is a giant, hybrid-morphology galaxy showing both an extended ellipsoidal stellar bulge,  girdled by an equatorial disc of  gas and dust. By any standards it has exceptional physical  properties which sets
it apart from most galaxies. It has some resemblance to
the nearby Sombrero galaxy (NGC 4594),  although UGC~12591 is much larger in size. 
The Keplerian disc rotation velocity of UGC~12591 is 
extremely high,  more than twice that of the rotational velocity of the Milky Way and one of the highest
known so far. Following  on the earlier results of \citet{1978AJ.....83.1026R}, \citet{1986ApJ...301L...7G} presented the 21 cm HI gas profile of UGC~12591 smoothed to an effective resolution of 25 \kms and reported  the 50$\%$ peak line width of $\rm W_{50} = 1014\pm10 \kms$ and maximum rotational velocity  (corrected for inclination and relativistic Doppler effect) of  $\rm v_{rot} = 506\pm 6 \kms$(488.4$\pm$12.5\kms; \citet[][]{2014A&A...570A..13M}). They also measured similar high rotation velocity extending  upto $\sim 20$ kpc from center, via $\rm H\alpha$ and [NII] lines of ionized gas. This corresponds to the interior dynamical masses  of ${\rm M_{dyn}} = 1.16\times10^{12}\rm \Msun$ and $2.9\times10^{12}\rm \Msun$ at 20 and 50 kpc respectively,  which are  extremely large and comparable to the most massive spiral galaxies. 
With XMM-Newton \citet{2012ApJ...755..107D} detected  a large, soft 
X-ray halo/corona emission out to $\sim 50 - 80$ kpc from the galaxy centre having  X-ray luminosity $\rm L_{x}(0.1-10 \kev)=3.9\times10^{40}$ \lum and  temperature of the halo gas of $\rm T_{halo} = 0.64\pm0.03~keV$, which showed that some fraction of  baryon mass resides in an extended,  
circum-galactic warm-hot gaseous halo. This is an extremely important
finding which has strong bearing on understanding the star formation history and baryon content of this and other massive  galaxies. 

Fig.~\ref{fig:HST} shows the optical color image made from HST WFC3 filters, where the existence of  prominent, stratified dust lanes and  the large extent of the galaxy's massive  stellar  bulge  can be seen clearly.
It is possible that these stratified dust lanes are spiral arms seen in projection. Another small satellite galaxy  WISEA J232529.77+283021.3 can be seen in  the picture.

For UGC~12591, excellent quality data of \textit{Hubble Space Telescope} (hereafter HST) has been taken and  here we present a   detailed photo-metric study  of  this galaxy using  three colour data in F606W, F814W $\&$ F160W bands, respectively. In each band, we have decomposed the galaxy into its bulge and disc components and estimated the structural parameters associated to them. 

We also present the best fitting spectral energy distribution (SED) of UGC~12591. The SED fitting has been done using two different software packages; (Sec.~\ref{sec:sed_MAGPHYS} and Sec.~\ref{sec:sed_CIGALE}) for cross checking and for obtaining a number of ISM and
stellar parameters. From these fits we estimate various key properties of the galaxy, such as its star formation rate, stellar mass, dust mass etc (Table~\ref{tab:cig_tab}, Table~\ref{tab: magphys_tab}). In this paper we demonstrate how this giant galaxy
possesses an extremely low star formation rate and a large repository of baryons and we  discuss the astrophysical implications of our findings.

The structure of this paper is as follows. In \S2. we describe the data, the data reduction strategy for the optical and near-IR observations along with SED fitting processes taking UV to FIR observations into account. In \S3. we present our discussion, and the summary of the study is outlined in \S4. We assume $\rm H_0$ = 72 km\, s$^{-1}$ Mpc$^{-1}$, $\rm \Omega_M$=0.26 and $\rm \Omega_{\Lambda}$=0.74,  translating to a scale of 0.452\,kpc\, arcsec$^{-1}$ at the red-shift $z=0.023$ of UGC~12591.

\begin{figure*}
    \centering
    \includegraphics[width=\textwidth]{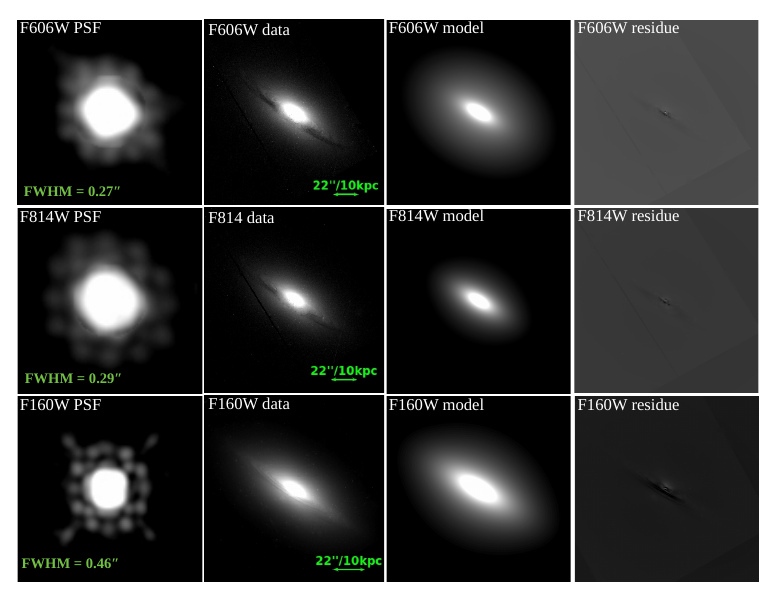}
    \caption{\textit{First Column:} The point spread functions (PSF) used in three HST WFC3 filters, [F606W (V band), F814W (I band) and F160W (H band, row wise], are shown with their corresponding FWHM.  \textit{Second Column:} Images of the  156\arcsec $\times$ 156\arcsec, 156\arcsec $\times$ 156\arcsec~ and 126\arcsec $\times$ 126\arcsec~ of  galaxy UGC~12591 in above  three HST filters are shown. \textit{Third and Fourth Columns:} The best fit models and their residuals for each bands are  shown, respectively.     While  fitting,  to reduce the background contamination from other sources (mainly stars and galaxies) and for robust fits, the  background has been masked in the IR F160W image and the  prominent  dust lanes have been masked carefully in the UVIS F606 and F814W images.}
    
    
    \label{fig:gal_res}
\end{figure*}

\section{Data Analysis} 
\label{sec:data}

\subsection{HST observations and data}
\begin{table}
    \centering
    \caption{Basic details of the HST data of UGC~12591.} 
    \label{tab:hst_data_details}
    \begin{tabular}{lccccr}
    \Xhline{2.7\arrayrulewidth}
    Filter & Band & Configuration & ExpTime & $ \lambda_{\rm central} $ & pixel size \\
           &      &               &  seconds    &   nm    & arcsec/pixel \\
    \hline
    F606W & V & WFC3/UVIS & 2619.00 & 588.8 & 0.039\\
    F814W & I & WFC3/UVIS & 2739.00 & 803.4 & 0.039\\
    F160W & H & WFC3/IR   & 5596.93 & 1536.9 & 0.090\\
    \hline
    \end{tabular}
   \end{table}
We have  used the  exquisite,   HST multi-band archival data (HST Observing Program 13370, PI: Georgiev I.Y.) for UGC~12591 which was observed  using the Wide Field Camera 3 (WFC3) in June 2014. 
WFC3 offers high resolution, sensitivity and wide field imaging capability over a wavelength range of near ultraviolet 200 nm to near infrared 1700 nm. WFC3 has two  imaging channels,  namely Ultraviolet-Visible (UVIS) and Infrared (IR). UVIS has a field of view of $\rm162\arcsec\times162\arcsec$ for the wavelength range of $\rm200-1000$ nm with pixel scale of $\rm 0.039\arcsec\times0.039\arcsec$. The IR channel has a field of view of $\rm 136\arcsec\times123\arcsec$ for the wavelength range of $\rm800-1700$ nm with a pixel scale of $\rm 0.13\arcsec\times0.12\arcsec$. WFC3 provides a choice of wide, medium and narrow band filters for both UVIS and IR. The UVIS and IR channel have 62 and 15 of these filters respectively with 1 grism in the UVIS and 2 in the IR channel. The UVIS detector operates in ACCUM mode in order to produce time integrated images. The IR detector  operates in MULTIACCUM mode, where accumulated signal can be read out non$-$destructively multiple times without causing harm to other pixels.


Deep observations were  taken  using age-metallicity sensitive, wide-band  V, I, and H filter sets  [V (F606W), I (F814W) and H (F160W)]  respectively. F606W and F814W have  rectangular  bandpass of width  218.2 nm and 153.6 nm respectively, whereas F160W has bandpass  of 268.3 nm. The peak system throughput of F606W, F814W and F160W are 0.29, 0.23 and 0.56 respectively. The background contamination has been significantly reduced by the high spatial resolution of  WFC3  channels  which is not achievable with ground based observations. The filter specific details about the HST data are tabulated in Table~\ref{tab:hst_data_details}.

We retrieved all of the FITS images from Hubble Legacy Archive\footnote{Based on observations made with the NASA/ESA Hubble Space Telescope, and obtained from the Hubble Legacy Archive, which is a collaboration between the Space Telescope Science Institute (STScI/NASA), the Space Telescope European Coordinating Facility (ST-ECF/ESA) and the Canadian Astronomy Data Centre (CADC/NRC/CSA)}.  


\begin{figure*}
    \includegraphics[scale=0.225]{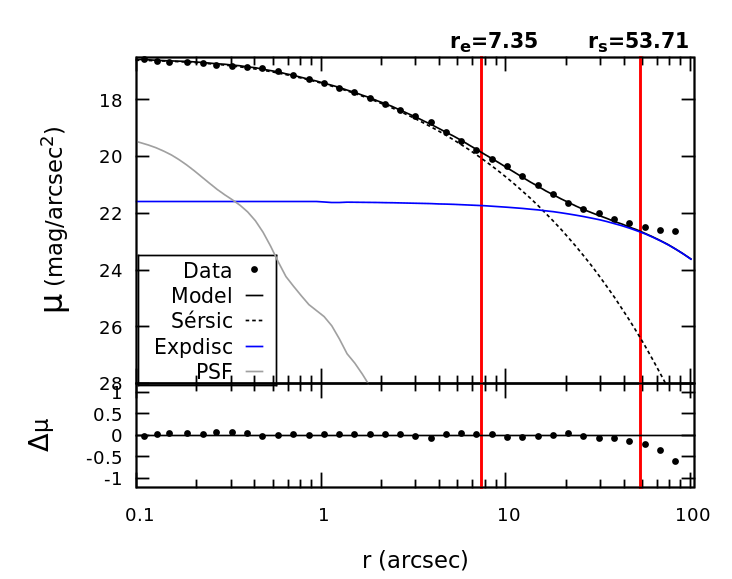}
    \includegraphics[scale=0.225]{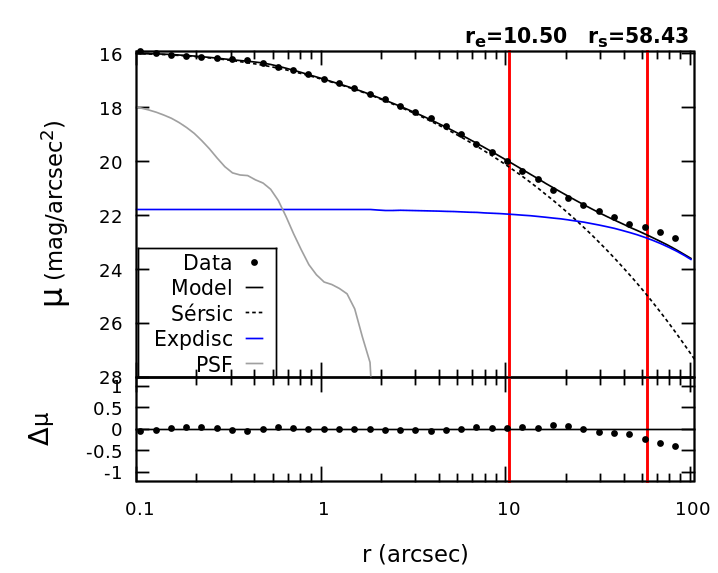}
    \includegraphics[scale=0.225]{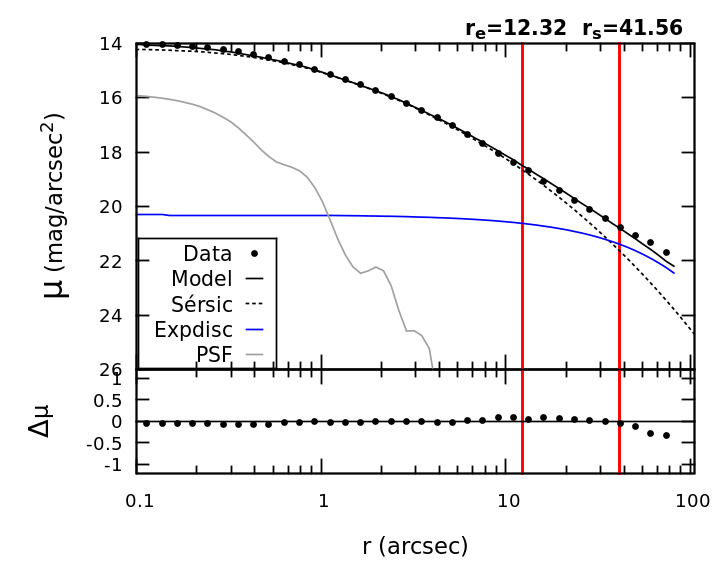}
    \caption{The left, middle and the right panels show  the model and observed surface 
    brightness profiles ($\rm \mu$ in mag$/$arcsec$^{2}$) in F606W,  F814W and F160W  bands respectively. The composite model (black line) fitted through observed  data (black dots) is decomposed
    into a \sersic~ profile, exponential disc profile and a PSF model
    which are shown in the legend.  In  bottom panels the residuals  ($\rm \Delta\mu$)  between the
    model and data have been shown. 
    In each plot the effective radius $\rm r_{e}$ corresponding to \sersic~ profile and the disc scale length $r_{s}$ corresponding to the Exponential-disc profile (see~Subsec.~\ref{subsec:bdd}) are shown  by vertical red lines. The best fit values of $\rm r_{e}$, $\rm r_{s}$ and other parameters are listed in Table~\ref{tab:gal_tab}. For $\rm r~\le~r_{e}$ the resultant model in each band is highly dominated by the \sersic~ function. And for $\rm r~\ge~r_{s}$ the model is dominated by exponential-disc function. It is  clearly seen from the plots for F606W (V) and F814W (I) that they have extended disc scale lengths.  The near infra-red  F160W (H) profile is  highly bulge dominated and its disc scale length is also less than that for F606W and F814W. For the region where $\rm r_{e}~\le~r~\le~r_{s}$, both the \sersic~ and the exponential functions  contribute to the model significantly.}
    \label{fig:sbp}
\end{figure*}

\begin{table}
    \centering
    \caption{Radiative flux  in different   passbands (effective wavelength $\rm \lambda_{eff}$) is  shown which is used for the SED fitting by {\fontfamily{qcr}\selectfont CIGALE and {\fontfamily{qcr}\selectfont GALFIT}}. The data is taken from $^{1}$\href{https://ned.ipac.caltech.edu/}{NED}
     $\And$ $^{2}$\href{http://cdsportal.u-strasbg.fr/}{CDS}. The references are: [a] \citet{2012AAS...21934001S} [b] \citet{2019AJ....158..138S}, [c] \citet{2015AAS...22533616H}, [d] \citet{2003yCat.2246....0C} [e] \citet{2013wise.rept....1C}, [f] \citet{1989ApJS...70..329K}.} 
    \label{tab:flux_data_ned}
    \begin{tabular}{lcrrc}
    \Xhline{2.7\arrayrulewidth}
    Passband & $\rm \lambda_{eff}$ & Flux & Flux Error & References \\
             & \micron             & mJy &  mJy        &\\
    \hline
    GALEX FUV & 0.1529 & 0.0307 & 0.00721 & 1[a]\\
    GALEX NUV & 0.2312 & 0.0795 & 0.00638  &1[a]\\
    SDSS u & 0.3519 & 1.90 & 0.01 &2[b]\\
    SDSS g$'$ & 0.4820 & 6.02 & 0.2893 &2[c]\\
    SDSS r$'$ & 0.6247 & 16.3 & 0.8 &2[c]\\
    SDSS i$'$ & 0.7635 & 27.2 & 1.7989 &2[c]\\
    2MASS J & 1.2390 & 62.9 & 0.232  &1[d]\\
    2MASS H & 1.6500 & 82.3 & 0.304  &1[d]\\
    2MASS Ks & 2.1640 & 73.9 & 0.273  &1[d]\\
    WISE W1 & 3.3500 & 41.9 & 0.232 &1[e]\\
    WISE W2 & 4.6000 & 22.8 & 0.126 &1[e]\\
    WISE W3 & 11.560 & 17.3 & 0.240 &1[e]\\
    WISE W4 & 22.090 & 11.3 & 1.40 &1[e]\\
    IRAS 60 & 60.000 & 230 & 34 &1[f]\\
    IRAS 100 & 100.00 & 1550 & 159 &1[f]\\
    \hline
    \end{tabular}
\end{table}

\subsection{SED input data}

Astronomical objects emit radiation across the electromagnetic spectrum. In spectral energy distribution (SED) fitting the radiative flux of any galaxy observed over UV to sub-mm is analysed to find out  several physical parameters like the star formation rate (SFR), stellar mass, dust mass etc. In order to fit the SED, we have made use of the available data  of  wide wavelength range,  spanning from GALEX in UV, SDSS in optical, 2MASS in near-IR, WISE in mid-IR and IRAS in the far-IR. The flux values from different telescopes as a function of central  effective wavelength have been tabulated in Table~\ref{tab:flux_data_ned}.


\subsection{Bulge-disc decomposition} 
\label{subsec:bdd}

\begin{table*}
    \centering
    \caption{Here the best fit parameters of {\fontfamily{qcr}\selectfont GALFIT} is shown in three filters V, I and H. The values in square brackets  are held fixed (priors) during the modeling in order to achieve better results. The \sersic, Exponential and PSF function have seven, six and three free parameters respectively. In the table we  exclude the central coordinates $\rm (x_{0}, y_{0})$ of these functions in 2D fit that are also free parameters as these will vary for different pixel dimensions of images taken to fit. The magnitudes given here are in AB scale.}
    \label{tab:gal_tab}
    \begin{tabular*}{\textwidth}{@{\extracolsep{\fill}}lccccccccccccc}
    \Xhline{2.7\arrayrulewidth}
    Filter&Band&\sersic~ Mag.&$\rm r_{e}$&n&\sersic~ $\rm \frac{b}{a}$&\sersic~ PA&Exp. Mag.&$\rm r_{s}$&Exp. $\rm \frac{b}{a}$&Exp. PA&PSF Mag.&$\rm N_{dof}$& $\rm \chi^{2}_{\nu}$ \\[4pt]
    -&-&mag&arcsec&-&-&deg&mag&arcsec&-&deg&mag&-&- \\
    \hline
    F606W&V&13.45&7.35&2.32&0.46&59.31&11.26&53.71&0.74&[59.00]&21.20&16007988&0.81\\
    F814W&I&12.70&10.50&3.01&0.46&58.74&11.12&58.43&0.83&54.12&19.65&16007987&0.99\\
    F160W&H&10.80&12.32&3.68&0.42&59.90&10.42&41.56&0.84&[59.00]&16.76&1962788&1.25\\
    \hline
    \end{tabular*}
\end{table*}

Extremely high spatial resolution offered by HST has some definite advantages
over ground-based imaging. For example,  galaxies with central bulge  and  strong 
light concentration,  a  surface brightness profile in 
poor  seeing conditions could give the illusion of a  central light deficit. 
Atmospheric blurring also affects the fitted  light profile creating a spurious  break/flattening on its slope. The  above effect is
more pronounced when the true light concentration is on a scale comparable to, or below the seeing  which can
mislead one to infer the existence of a depleted core, which results  
due a dynamical influence of a  super massive black hole (SMBH). 
HST can largely mitigate this problem.
Moreover, the  simultaneous  fit  of  wide-wavelength range, high resolution HST
images makes the bulge-disc  decomposition more robust against such  spurious effects.

The  structural decomposition  was done using {\fontfamily{qcr}\selectfont GALFIT} software \citep[][]{2002AJ....124..266P} on the HST F606W, F814W and F160W band images. We first mask the foreground stars in  the images. UGC 12591 contains thick dust lanes, primarily dominant in the UVIS bands (Fig.~\ref{fig:gal_res}). Masking these dust lanes is necessary to avoid its effects on structural parameters obtained from the 2D fit. The masking was done using the {\fontfamily{qcr}\selectfont IMEDIT} task in {\fontfamily{qcr}\selectfont IRAF} \citep{1986SPIE..627..733T, 1993ASPC...52..173T}. After masking for unwanted sources and dust lanes for each of the filters  the 2D decomposition is performed. We use the Point Spread Function (PSF) obtained from STScI for convolution while fitting, and used $\rm 1401\times1401$ pixel fitting region for F160W and $\rm 4001\times4001$ pixel region for both of F606W and F814W images centered on the brightest pixel of the object. To fit the image we specify the profile function as well as the initial estimates for the model parameters. For \sersic~ model these parameters are; position of the centre of the profile, integrated magnitude ($\rm mag$), effective radius ($\rm r_{e}$), \sersic~ index $\rm (n)$, axis ratio $\rm (b/a)$ and the position angle $\rm (PA)$. The best-fit model is produced through multiple iteration based on these initial parameters and then convoluted with the corresponding WFC3 PSF. The best-fit is based on $\chi^{2}_{\nu}$ minimization using the Levenberg-Marquardt downhill-gradient method. We have used the sigma image generated internally by {\fontfamily{qcr}\selectfont GALFIT} in this fit. Although we used
the pipeline, we vary the initial conditions of different parameters to check whether the final results correspond to a global
minimum. We found that the results are stable against different
initial conditions. 

Considering the disc morphology and a prominent bulge in UGC 12591, initially we model the surface brightness profile with two components, viz. \sersic~ model for the bulge and an exponential disc function. The \sersic~ profile  is defined as,
\begin{equation}\label{eq:sersic}
    \rm{\Sigma(r) = \Sigma_{e} \, exp \left[-\kappa \, \left(\left(\frac{r}{r_{e}}\right)^{1/n} - 1\right)\right]}
\end{equation}
Here, $\rm \Sigma(r)$ and $\rm \Sigma_{e}$ are pixel surface brightness at radius $\rm r$ and $\rm r_{e}$ respectively, $\rm r_{e} $ is effective (half-light) radius, $\rm n $ is the \sersic~ index, and $\rm \kappa $ is a dependent variable coupled to $\rm n$ such that $\rm r_{e}$ encloses half of total brightness. 

The exponential-disc function is defined as,
\begin{equation}\label{eq:expdisk}
    \rm{\Sigma(r) = \Sigma_{0} \, exp \left(-\frac{r}{r_{s}} \right)}
\end{equation}
Here, $\rm \Sigma_{0} = \Sigma(r)|_{r=0}$ and $\rm r_{s} $ is disc the scale length. 
The two component  model fits well the bulge and the disc, however the residual images show an unresolved light excess at the centre of the galaxy, probably attributed to an AGN activity and/or  a  compact nuclear star cluster. To accommodate this excess light we added  an additional  PSF function along with the \sersic~ and exponential disc components. The results show a significant improvement of the residuals and  model parameters. The final best fit results are given  in Table~\ref{tab:gal_tab} showing excellent  reduced $\chi^{2}_{\nu}$  and depicted in  Fig.~\ref{fig:gal_res}.

\begin{figure*}
	\includegraphics[scale=0.58]{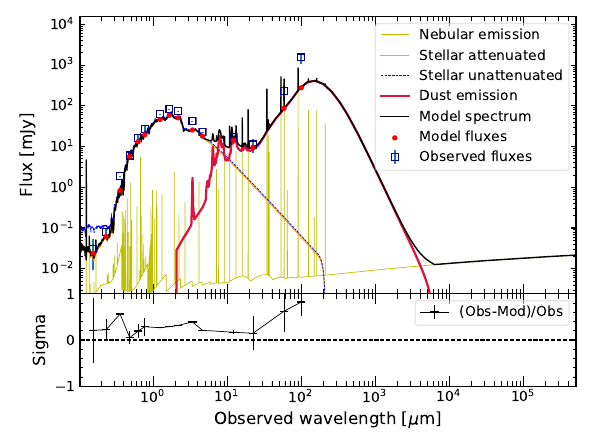}
    \includegraphics[scale=0.58]{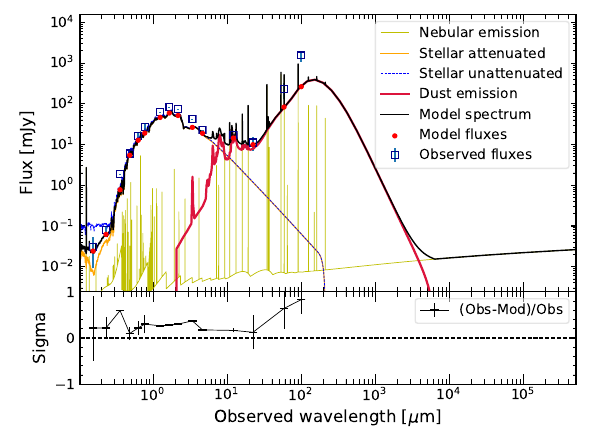}
    \includegraphics[scale=0.58]{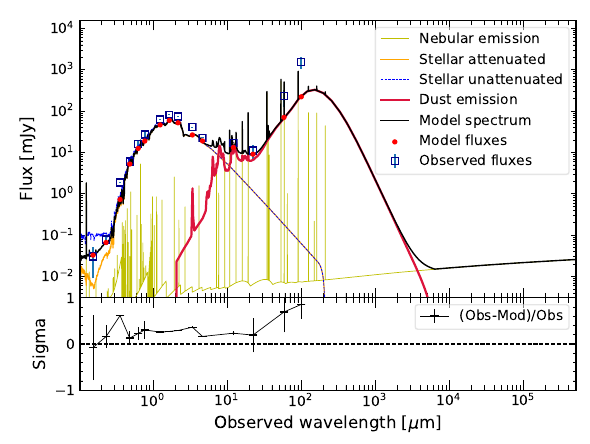}
    \caption{Here the Spectral energy distribution fitting of UGC~12591 using {\fontfamily{qcr}\selectfont CIGALE} has been presented for different AGN fractions. \textit{Left:} AGN fraction = 0, \textit{Middle:} AGN fraction = 0.05, \textit{Right:} AGN fraction = 0.10. Observed flux  over the wavelength range of $\rm 0.153 - 100~\micron$ has been fitted to estimate  some key parameters of galaxy     shown in Table~\ref{tab:cig_tab}.
    Here sigma represents the relative residual flux. The blue squares are the input data points.  The red dots are the best fit fluxes from SED fitting. 
    Different ISM component contributing to SED fluxes  are shown in the figure legends. No signature of strong  radiative feedback from an AGN could be detected in the SED models.
    }
    \label{fig:sed_fit}
\end{figure*}

\subsection{Surface brightness profiles of UGC~12591} 
\label{subsec:sbp}

The visualization of how good the modeling has been can be depicted in 1D radial profile plots even though the non-axisymmetric features of the surface brightness profiles are lost in this representation. After getting the models generated by {\fontfamily{qcr}\selectfont GALFIT} (Fig.~\ref{fig:gal_res}), the {\fontfamily{qcr}\selectfont ellipse} routine of {\fontfamily{qcr}\selectfont PyRAF/STSDAS} was used  to plot the surface brightness ($\rm \mu$) variation as a function of the semi-major axis distance ($\rm r$) of the galaxy by fitting concentric elliptical isophotes  of constant surface brightness. The procedure is applied to the final PSF convolved model as well as to the models corresponding to the sub components, such as the \sersic, exponential-disc and PSF, in each of the three bands to calculate  individual contribution of the sub components to the composite light profile of the galaxy. The results are  shown in the Fig.~\ref{fig:sbp}. The goodness of the fit is quantified  by the
 residual $\rm \Delta\mu$  shown, that indicates the difference between data and corresponding model. The contribution of the \sersic, exponential-disc  and PSF profiles can be seen in each of the bands.

\label{sec:sed_MAGPHYS}
\begin{figure*}
	\includegraphics[width=\textwidth]{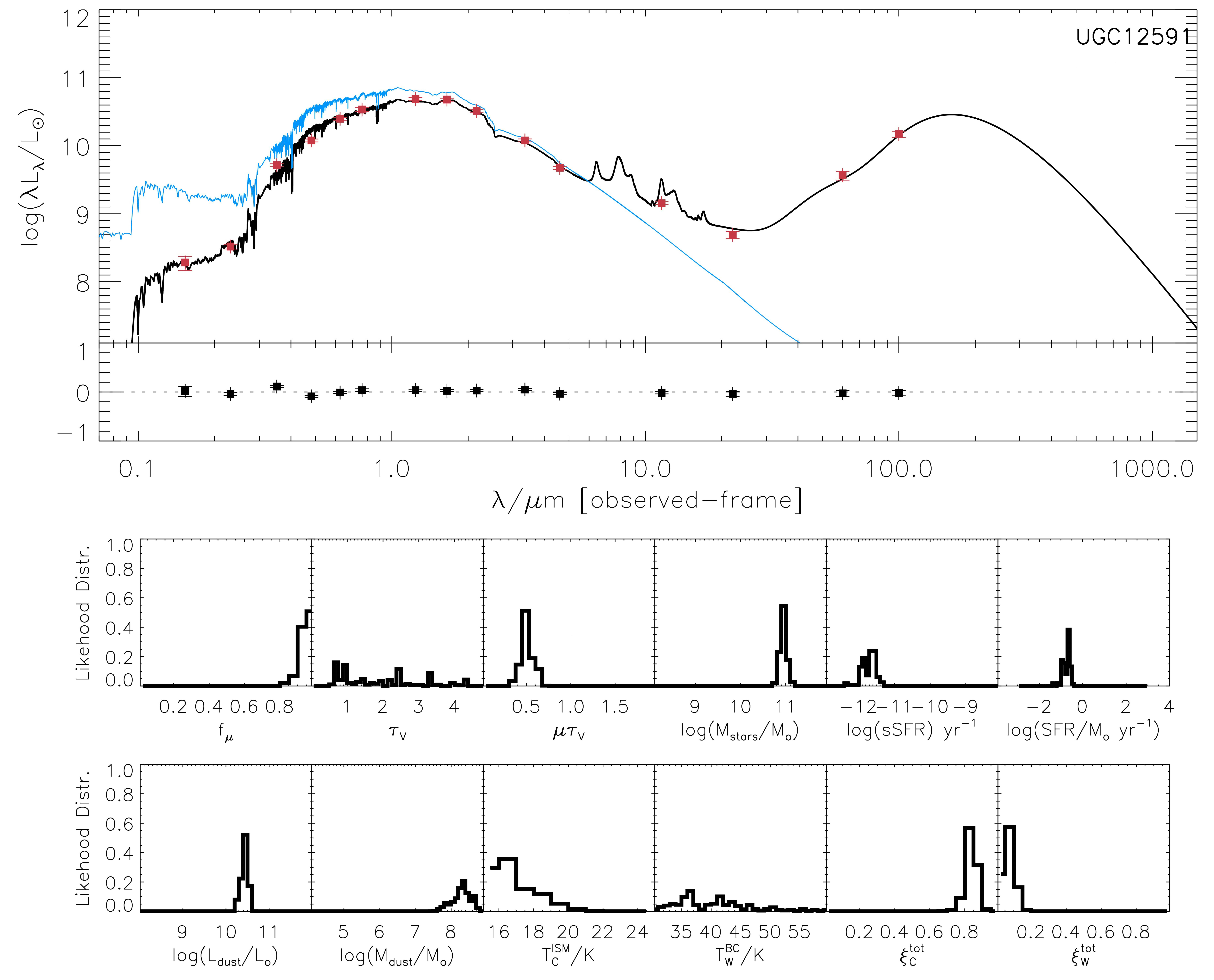}
    \caption{Result of  SED fitting with  {\fontfamily{qcr}\selectfont MAGPHYS} is shown here. The black solid line is the best-fit model to the observed SED with $\rm \chi^{2} = 3.96$, and observed flux data points are shown  in red.  The data points are composed of two GALEX, four SDSS, three 2MASS, two IRAS, and four WISE bands.  The blue solid line shows the unabsorbed stellar  spectrum. The bottom panel shows the normalised residuals. The lower part of the figure gives the likelihood distribution of best fit physical parameters for UGC~12591.
    These are;
    (1) fraction of total luminosity contributed by dust in the ambient (diffuse) ISM, $f_\mu$ (2) average V band dust attenuation, $\rm \tau_{V}$ 
(3) Total  effective V-band  absorption  optical  depth  of  the  dust  seen  by  young  stars inside birth clouds $\rm \mu\tau_V$,
(4) Stellar Mass, $ \rm log(M_{\star}/\Msun)$, 
(5) specific star formation rate, $ \rm log(sSFR) = log(SFR/M_{\star}) \, yr^{-1}$,
(6) star formation rate, $\rm log(SFR/\Msun yr^{-1})$,
(7) total dust luminosity, $\rm log(L_{dust}/L_{\odot})$,
(8) total dust mass, $\rm  log(M_{dust}/\Msun)$, 
(9) equilibrium temperature of cold dust in the ambient ISM,  $\rm T_{C}^{ISM}/K$,
(10) equilibrium temperature of warm dust inside the birth clouds, $ \rm T_{W}^{BC}/K$,  (11) Total contribution by cold dust in the infrared emission, $\rm \xi_C^{tot}$ and, 
(12)  Total contribution by warm dust in  the infrared emission, $\rm \xi_W^{tot}$.}  
\label{fig:sed_magphys}
\end{figure*}


\subsection{SED fitting with {\fontfamily{qcr}\selectfont CIGALE}} 
\label{sec:sed_CIGALE}

\begin{table*}
    \centering
    \setlength{\tabcolsep}{2.5pt}
    \caption{Here the best fit model parameters of the SED fitting by {\fontfamily{qcr}\selectfont CIGALE} is shown for different AGN fraction. The parameters mentioned from left to right are as follows: AGN $-$ fraction that indicates contribution  of an active galactic nuclei, $\rm M^{old}_{\star} -$ mass of older stellar population, $\rm M^{young}_{\star}-$ mass of younger stellar population, $\rm M_{\star} -$ total stellar mass, $\rm M^{old}_{gas} -$ ISM gas mass of old stars gas, $\rm M^{young}_{gas} -$ ISM gas mass of young stars, $\rm M_{gas} -$ total gas mass, $\rm SFR -$ instantaneous SFR, $\rm SFR_{10} -$ SFR averaged over 10 Myrs, $\rm SFR_{100} -$ SFR averaged over 100 Myrs, sSFR $-$ specific star formation rate, $\rm L_{\star} -$ stellar luminosity, $\rm L_{dust} -$ dust luminosity, $\rm \chi^{2}_{\nu} -$ reduced chi square. Note that, as {\fontfamily{qcr}\selectfont CIGALE} does not provide the uncertainties related to the best fit values, we here mention the bayesian uncertainties as the measurements of error.}
    \label{tab:cig_tab}
    \begin{tabular*}{\textwidth}{@{\extracolsep{\fill}}lccccccccccccr}
    \Xhline{2.7\arrayrulewidth}
    AGN & $\rm M^{old}_{\star}$ & $\rm M^{young}_{\star}$ & $\rm M_{\star}$&$\rm M^{old}_{gas}$&$\rm M^{young}_{gas}$&$\rm M_{gas}$&$\rm SFR$&$\rm SFR_{10}$&$\rm SFR_{100}$&sSFR&$\rm L_{\star}$&$\rm L_{dust}$& $\rm \chi^{2}_{\nu}$ \\[4pt]
    fraction&$\rm 10^{11} \Msun$&$\rm 10^{5} \Msun$&$\rm 10^{11} \Msun$&$\rm 10^{10} \Msun$&$\rm 10^{4} \Msun$&$\rm 10^{10} \Msun$&$\rm \Msun yr^{-1}$&$\rm \Msun yr^{-1}$&$\rm \Msun yr^{-1}$&$\rm 10^{-13} yr^{-1}$&$\rm 10^{37} W$&$\rm 10^{36} W$& \\
    \hline
    0.00&1.6$\pm$0.1&10.0$\pm$0.7&1.6$\pm$0.1&7.1$\pm$0.5&4.3$\pm$0.3&7.1$\pm$0.5&0.105$\pm$0.007&0.105$\pm$0.007&0.110$\pm$0.007&6.6$\pm$0.6&2.2$\pm$0.1&2.0$\pm$0.2& 4.74 \\
    0.05&1.6$\pm$0.1&10.0$\pm$0.7&1.6$\pm$0.1&7.0$\pm$0.5&4.2$\pm$0.3&7.0$\pm$0.5&0.104$\pm$0.007&0.104$\pm$0.007&0.109$\pm$0.007&6.5$\pm$0.6&2.1$\pm$0.1&1.9$\pm$0.2& 4.89 \\
    0.10&1.5$\pm$0.1&9.9$\pm$0.7&1.5$\pm$0.1&6.9$\pm$0.5&4.2$\pm$0.3&6.9$\pm$0.5&0.102$\pm$0.007&0.103$\pm$0.007&0.107$\pm$0.007&6.8$\pm$0.6&2.1$\pm$0.1&1.6$\pm$0.1& 5.46 \\
    \hline
    \end{tabular*}
\end{table*}

\begin{table*}
    \centering
    \caption{Here, the model parameters used in {\fontfamily{qcr}\selectfont CIGALE} are shown. Note that the parameters that are mentioned here with one value are passed through multiple trials to find out those best values for which we get the best results. We use the default values for the parameters that are not listed here.}
    \label{tab:cigale_model}
    \begin{tabular*}{\textwidth}{@{\extracolsep{\fill}}lccl}
    \Xhline{2.7\arrayrulewidth}
    Module & Parameter & Values & References \\[0.5ex]
    \hline
    Star formation history & E-folding time of main stellar population model & 1 Gyr & {\cite{2019A&A...622A.103B}}\\
    (\textbf{sfhdelayed})& Age of main stellar population & 10 Gyr & \\
    & E-folding time of late starburst population model & 40 Myr & \\
    & Age of the late burst & 25 Myr & \\[0.5ex]
    \hline
    Stellar population & Initial mass function & Salpeter & {\cite{2003MNRAS.344.1000B}}\\
    (\textbf{bc03})& Metallicity & 0.05 &\\[0.5ex]
    \hline
    Dust attenuation & V-band attenuation of young population & 0.7, 0.8, 0.9, 1 &{\cite{2019A&A...622A.103B}}\\
    (\textbf{dustatt\_powerlaw})& Reduction factor of $\rm A_{v}$ for old population compared to young one& 0.0, 0.11, 0.55 &\\
    & UV bump wavelength & 120, 150, 217.5 nm &\\
    & UV bump width & 50, 60, 70, 80 nm &\\
    & UV bump amplitude & 11, 12, 15, 18 &\\
    & Slope for power law continuum & -1.0, -0.5, 0.0, 1.0 &\\[0.5ex]
    \hline
    Dust emission & AGN fraction & 0.0/0.5/0.10 &{\cite{2014ApJ...784...83D}}\\
    (\textbf{dale2014})& $\rm \alpha$ & 1.0, 2.0, 2.5, 3.0 &\\[0.5ex]
    \hline
    \end{tabular*}
\end{table*}


The SED fitting is done with the python based software {\fontfamily{qcr}\selectfont CIGALE} (Code Investigating Galaxy Evolution) \citep{2019A&A...622A.103B} which extends the SED fitting algorithm written by \citet[][]{2005MNRAS.360.1413B, 2009A&A...507.1793N}. We make use of all the available flux in literature over the wavelength ($\rm 0.1 - 100 \micron$) range of far-ultraviolet to far-infrared (Table~\ref{tab:flux_data_ned}) and implement modules to compute the best fit SED which are: (1) \textit{sfhdelayed} for star formation history \citep[][]{2019A&A...622A.103B} (2) \textit{bc03} for stellar population \citep[][]{2003MNRAS.344.1000B} (3) \textit{nebular} for nebular emission \citep[][]{2011MNRAS.415.2920I} (4) \textit{dustatt$\_$powerlaw} for dust attenuation \citep[][]{2019A&A...622A.103B} (5) \textit{dale2014} \citep[][]{2014ApJ...784...83D} for dust emission and (6) \textit{redshifting} \citep[][]{2006MNRAS.365..807M}. The SED fitting done by {\fontfamily{qcr}\selectfont CIGALE} is shown in Fig.~\ref{fig:sed_fit}. The model outputs consist of two peaks, one for stellar emission indicated by the blue dotted-line and one for the dust emissions indicated by red continuous line.  The greenish-yellow lines designate the Nebular emission of ionized gas at different wavelengths. 

We have also considered three SEDs modelled  with different AGN fractions of $\rm 0\%,  5\% $ and $ \sim 10\%$. It is defined as the AGN contribution to the total IR luminosity from $\sim 5 - 1000 \mu m$. AGN fraction explicitly
takes into account three  emission components through a radiative transfer model: the primary source surrounded by a dusty torus, the scattered
emission by dust, and the reprocessed thermal dust emission \citep[][]{2019A&A...622A.103B}. 
 It can be seen from the SED fit parameters (Table~\ref{tab:cig_tab}) that as the AGN fraction is increased the $\rm \chi_{\nu}^{2}$ increases but the best fit parameters are not drastically changed. Thus we conclude that
 at present a powerful  AGN  (hosted by a 
supermassive black hole) is not contributing much to the  mid-IR  bands,  ruling out the presence of strong
radiative feedback from a bolometrically  bright AGN. NVSS has
detected a  compact central radio source  associated with UGC~12591 having luminosity of $L_{\rm 1.4GHz} = $ 3.01$\pm 0.92 \times 10^{32}$ W \citep{1998AJ....115.1693C,2002AJ....124..675C} which can  possibly be attributed to radio emission of an AGN, although no large scale radio jets have been detected.

The results of SED fitting by {\fontfamily{qcr}\selectfont CIGALE} is shown in Table~\ref{tab:cig_tab}. Here we only mention the best fit parameters that are important for our analysis: (1) old stellar mass $\rm M_{\star}^{old}~(\Msun)$ (2) young stellar mass $\rm M_{\star}^{young}~(\Msun)$ (3) total stellar mass $\rm M_{\star}~(\Msun)$ (4) ISM gas mass of old stars $\rm M_{gas}^{old}~(\Msun)$ (5) ISM gas mass of young stars $\rm M_{gas}^{young}~(\Msun)$ (6) total ISM gas mass $\rm M_{gas}~(\Msun)$ (7) instantaneous star formation rate SFR $(\Msun yr^{-1})$ and specific star formation rate $\rm sSFR =  SFR/M_{\star} \, (yr^{-1})$ (8) SFR averaged over 10 and 100 Myrs (9) stellar Luminosity $\rm L_{\star}~(Watt)$ (10) dust luminosity $\rm L_{dust}~(Watt)$.

\subsection{SED fitting with {\fontfamily{qcr}\selectfont MAGPHYS}} 

We also fitted the SED with the {\fontfamily{qcr}\selectfont MAGPHYS} package (Multi-wavelength Analysis of Galaxy Physical
Properties\, \citep{2008MNRAS.388.1595D}). This package allows to fit the whole SED, from the UV to the far-IR, by relating the optical and IR libraries in a physically consistent way. For the SED fitting we used 15 photometeric bands, i.e. from the GALAX NUV to the IRAS 100 $\micron$ and computed the SED with the stellar-population synthesis models of \cite{2003MNRAS.344.1000B}. The  wavelengths  and flux values used for the SED fitting are listed in Table.~\ref{tab:flux_data_ned}. The best fit SEDs were obtained by fixing the redshift to {\it z} = 0.023. The best fit output from the {\fontfamily{qcr}\selectfont MAGPHYS} package is shown in Table~\ref{tab: magphys_tab} and Fig.~\ref{fig:sed_magphys}. In this figure the black solid line is the best-fit model to the observed SED and available data points are 
shown in red.  
The blue solid line shows the unabsorbed stellar  spectrum. The bottom panel shows the residuals ($L_{obs,\lambda}$ - $L_{model,\lambda}$ / $L_{obs,\lambda}$). The lower part of the figure gives the likelihood distribution of  several best fit physical parameters for UGC~12591.  They are;  (1) fraction of total dust luminosity contributed by dust in the ambient (diffuse) ISM $f_\mu$ (2) average V band dust attenuation, $\rm \tau_{V}$
(3) Total  effective V-band  absorption  optical  depth  of  the  dust  seen  by  young  stars inside birth clouds $\rm \mu\tau_V$ 
(4) stellar Mass, $ \rm log(M_{*}/\Msun)$ 
(5) specific star formation rate, $ \rm log(sSFR) \, yr^{-1}$
(6) star formation rate, $\rm log(SFR/\Msun\, yr^{-1})$ 
(7) total dust luminosity, $\rm log(L_{dust}/L_{\odot})$  
(8) total dust mass, $\rm  log(M_{dust}/\Msun)$ 
(9) equilibrium temperature of cold dust in the ambient ISM,  $\rm T_{C}^{ISM}/K$
(10) fractional contribution by cold dust to the dust luminosity of the ambient ISM, $ \rm T_{W}^{BC}/K$  (11) Total contribution by cold dust in the infrared emission, $\rm \xi_C^{tot}$ and
(12)  Total contribution by warm dust in to the infrared emission, $\rm \xi_W^{tot}$.

The likelihood distributions of each of the above parameters are shown in Fig.~\ref{fig:sed_magphys}. We take the median value of the distribution as the measurement of a parameter with 16th to 84th percentile range to be the $\rm 1\sigma$ confidence level (Table~\ref{tab: magphys_tab}).

%


\begin{table*}
    \centering
    \setlength\extrarowheight{0.8pt}
    \setlength{\tabcolsep}{5.2pt}
    \caption{SED fitting parameters of UGC~12591 using {\fontfamily{qcr}\selectfont MAGPHYS}. }
    \label{tab: magphys_tab}
    \begin{tabular}{lccccccccccr}
    \Xhline{2.7\arrayrulewidth}
         sSFR & $\rm M_{\star}$ & $\rm L_{dust}$ & $\rm M_{dust}$ & SFR & $\tau_{V}$ & $\mu \tau_{V}$ & $f_{\mu}$ & $\rm T_{W}^{BC}$ & $\rm T_{C}^{ISM}$ & ${\rm \xi_{C}^{tot}}$ & ${\rm \xi_{W}^{tot}}$ \\[4pt]
         $\rm 10^{-12}yr^{-1}$ & $\rm10^{10}\Msun$ & $\rm10^{10}\Lsun$ & $\rm 10^{8} \Msun$ & ${\rm M_{\odot} yr^{-1}}$ &  & & & K & K & & \\
         \hline
         $2.14^{+0.88}_{-0.94}$ & $9.12^{+1.38}_{-1.54}$ & $2.75^{+0.48}_{-0.41}$ & $2.07^{+1.41}_{-1.05}$ & $0.210^{+0.037}_{-0.097}$ & $1.77^{+1.51}_{-1.06}$ & 
         $0.50^{+0.09}_{-0.07}$ & 
         $0.94^{+0.03}_{-0.02}$ & $39.49^{+6.90}_{-4.90}$ & $16.64^{+1.67}_{-1.10}$ & $0.84^{+0.03}_{-0.03}$ & $0.08^{+0.02}_{-0.04}$ \\
         \hline
    \end{tabular}
\end{table*}


\section{Discussion } 
\label{sec:disc_and_conc}

\subsection{Bulge-Disc structure of UGC~12591}
\label{sec:structure_ugc12591}

The morphology  of UGC~12591 has been decomposed into two principal   components of \sersic~bulge and exponential disc. The surface brightness profiles of the galaxy are presented in Fig.~\ref{fig:sbp} where $\rm r_{e}$ and  $\rm r_{s}$ are the scaling parameters of the bulge and disc components. It can be seen that the bulge effective radius $\rm r_{e}$ (\S~\ref{subsec:bdd}) ranges from $3.3 - 5.6$ kpc (Table~\ref{tab:gal_tab}) which indicates the extent of the bulge of the galaxy. From  V to H  band the value of $\rm r_{e}$ also increases, i.e. bulge is  more  prominent at longer 
wavelengths due to  dominance of old stellar  population in bulge.

On the other hand, disc scale length $\rm r_{s}$ (Eq.~\ref{eq:expdisk}) is around $24.3 - 26.4$ kpc in  V  and I  which decreases to $18.8$ kpc in H band. \citet{1996A&AS..117..393B} finds that the wavelength dependence of the scale lengths is due to the dust extinction. \citet{2003ApJ...582..689M} shows that the disc scale length decreases at longer wavelength due to higher concentration of older stars and dust in the central region relative to the outer disc region where the probability of star formation is higher \citep{2004A&A...415...63M}. The ratio of bulge effective radius to disc scale length ($\rm r_{e}/r_{s}$) for V, I and H band are given by approximately 0.14, 0.18 and 0.30 respectively, showing increment of this ratio as band wavelength increases. 

\sersic~ index $\rm n$   for UGC~12591 shows the same pattern as the effective radius $\rm r_{e}$. Bulge  $\rm n$ increases from $2.32$ to $3.68$ (Table~\ref{tab:gal_tab}) in  V to H bands. This falls in the range of  classical bulges which  have a much more centrally
peaked light profile, contain a higher fraction of total light,
and their \sersic~ index is larger (\rm $n \approx 2 - 6$) 
than pseudo-bulges, which have $\rm n < 2$ \citep[][]{2008AJ....136..773F}. The ratio of
rotation velocity to central velocity dispersion $\rm v_{rot}/\sigma_{\star} \sim 1.7$ is
high for a  bulge dominated system. Thus,  along with bulge dispersion support, rotation support plays a major role in the dynamics of this galaxy.

Looking at  the bulge (\sersic~)  magnitudes it can be seen that the bulge  rapidly becomes  brighter  from V to H as $\rm n$ increases.  This proves that  UGC~12591 indeed is a bulge dominated disc galaxy as is obvious   from Fig.~\ref{fig:sbp}. \citet{2004A&A...415...63M} showed that the bulges in early-type spirals are brighter, similar to elliptical galaxies, and larger than that of the late-type galaxies for which the bulges are comparable to their discs. The possible reason is that preeminent  bulges of early-type spirals could have formed due to early mergers, which 
possibly explains  structural properties of  UGC~12591.

\subsection{The bulge-to-disc mass and luminosity ratios}

In Table~\ref{bulge-disc} we show the bulge and 
disc parameters of UGC~12591 derived using {\fontfamily{qcr}\selectfont GALFIT} in three HST colours. Mass and luminosity
of bulge and disc components are denoted by M or L with indicative subscripts. 
The B/D and B/T parameters represent the luminosity ratios of bulge to disc and bulge to total (bulge+disc) components respectively. Both these parameters are strongly colour dependent.
The mass
and luminosity of bulge component is largely dominated by old stellar population 
contributing  more to the H band NIR flux than in optical  V band. 

\begin{table}
    \centering
    \caption{Here, we show the  bulge and disc  mass (M) and luminosity (L) parameters of UGC~12591 derived using {\fontfamily{qcr}\selectfont GALFIT} best fit magnitudes shown in Table~\ref{tab:gal_tab}. We have used \citet[][]{2001ApJ...550..212B} to calculate the mass to light ratios, which are 1.78  (V-band), 1.34 (I-band) and 0.68 (H-band). The B/D and B/T represent the luminosity ratios of bulge to disc and bulge to total (bulge+disc) components respectively.}
    \label{tab:luminosity_mass}
    \begin{tabular*}{\columnwidth}{@{\extracolsep{\fill}}lcccccc}
    \Xhline{2.7\arrayrulewidth}
    Band & $\rm L_{bulge}$ & $\rm L_{disc}$ & $\rm M_{bulge}$ & $\rm M_{disc}$ & $\rm \frac{B}{D}$ & $\rm \frac{B}{T}$ \\[4pt]
      & $\rm 10^{10} \Lsun$ & $\rm 10^{11} \Lsun$ & $\rm 10^{10} \Msun$ & $\rm 10^{11} \Msun$ &  &  \\
      \hline
      V & 3.1 & 2.3 & 5.5 & 4.1 & 0.13 & 0.12 \\
      I & 5.1 & 2.2 & 6.8 & 2.9 & 0.23 & 0.19 \\
      H & 31 & 4.5 & 21 & 3.1 & 0.69 & 0.41 \\
      \hline
    \end{tabular*}
    \label{bulge-disc}
\end{table}

\subsection{Interpreting dust mass and dust emission}

UGC~12591 possesses  a prominent disc of  interstellar  dust and
cold, neutral HI gas in the equatorial plane.  Based on our SED fitting 
(Table~\ref{tab: magphys_tab}) we  highlight 
the properties of  the  dust component here.  
\citet[][]{2012MNRAS.419.2545R} did a study on dust properties and star formation history of the local sub millimetre detected galaxies, with some of them being early-type and passive spirals. The passive spirals are defined as early-type galaxies with dust abundance and low SFR. They   found  the  mean $\rm \log M_{dust} = 7.74 \Msun$ with $\rm M_{dust}$ in the range $\rm 10^{5} - 10^{8} \Msun$ for early-type galaxies,  and for passive spirals mean $\rm \log M_{dust} = 7.47$ with $\rm M_{dust}$ in the range $\rm 10^{6} - 10^{8}\Msun$. From SED fitting we find $\rm \log M_{dust} = 8.3 \Msun$ (Table~\ref{tab: magphys_tab} and Fig.~\ref{fig:sed_magphys})  consistent with the results of \citet[][]{2012MNRAS.419.2545R},  with dust mass for UGC~12591 falling in the highest end of  passive spirals. The ratio $\rm  M_{dust}/M_{\star} \approx 0.002$ is typical
of  passive galaxies \citep[][]{2017A&A...602A..68O}.

The estimated equilibrium temperature of cold dust in the ambient ISM is $\rm T_{C}  \approx 17 K$  and warm dust temperature $\rm T_{W} \approx 40 K$. In our SED fitting  a single black body mean temperature is assumed for the cold  dust. More likely,  a continuous range of temperatures, heated  by emission sources is needed \citep[][]{2015MNRAS.448..135B}. We find $f_{\mu}$, the fraction of total infra-red dust luminosity contributed by the ISM dust is as high as $\rm 0.94$ with total dust luminosity $\rm L_{dust} = (0.50-2.75) \times 10^{10} \Lsun$ (Table~\ref{tab:cig_tab} and Table~\ref{tab: magphys_tab}), much higher than the mean $\rm f_{\mu} = 0.58 $ for normal $\rm L_{\star}$ spiral galaxies  in \citet[][]{2012MNRAS.419.2545R}. However in  $\rm L_{dust}$, the  cold  dust contribution is much higher,  $\rm > 80 \%$ compared to warm dust $\rm <10 \%$ (Table~\ref{tab: magphys_tab} and Fig.~\ref{fig:sed_magphys}). This is  broadly consistent with the findings of  \citet{2017A&A...602A..68O} and \citet{2010MNRAS.403.1894D} that, the $\rm L_{dust} - M_{dust}$ plane placement of a galaxy is most sensitive to the temperature of the cold component, rather than the values of  $\rm L_{dust}$ or $\rm M_{dust}$.  This shows that  UGC~12591 is a passive galaxy in which the ISM  dust luminosity is dominated by heating from cooler, older stars, rather than hot, newborn stars. This is further confirmed by calculations of  star formation rate presented below. Future observations in mm/radio bands (mainly CO lines) and sub-mm/IR bands will better constrain the molecular gas and dust properties of this galaxy respectively.

\subsection{Star formation} 
\label{subsec:sfr}


\begin{figure*}
    \centering
     \includegraphics[scale = 0.35]{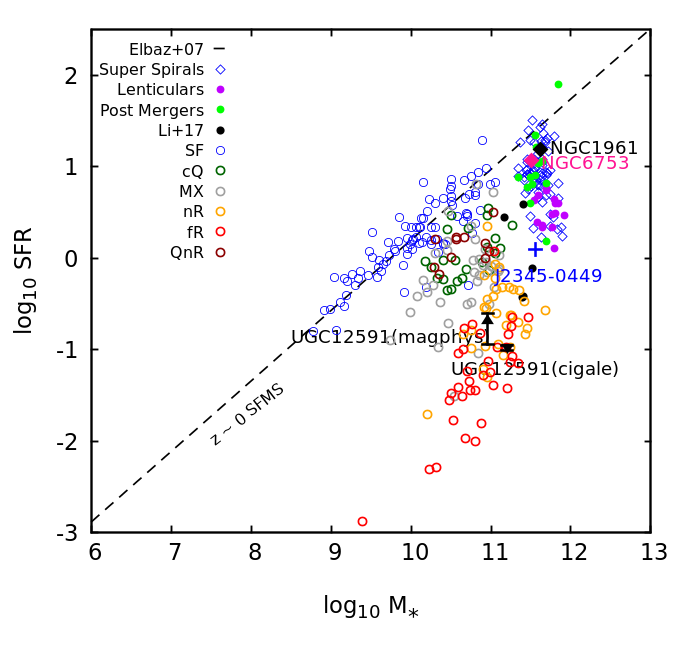}
     \includegraphics[scale = 0.35]{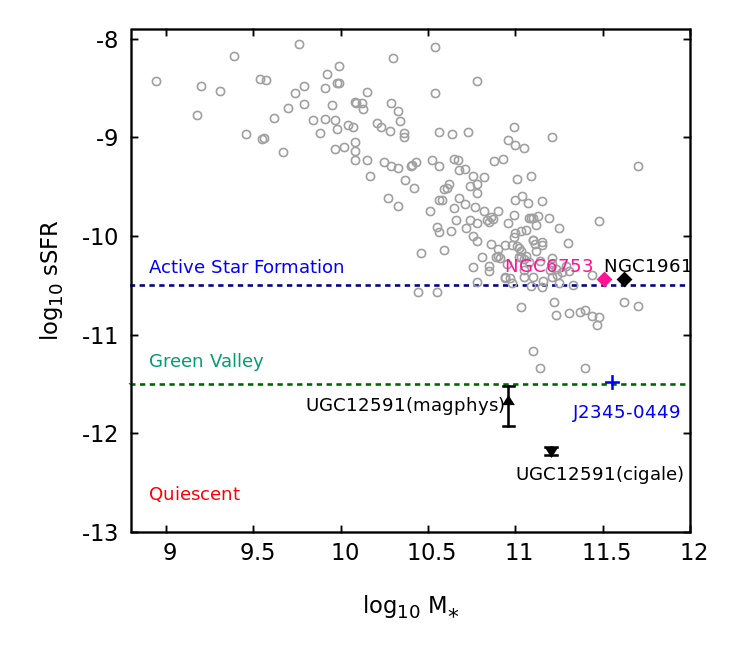}
     \caption{{Left:} Here, the variation of  SFR (\rm $\Msun yr^{-1}$) with stellar mass $\rm M_{\star}$ ($\rm \Msun$) has been shown for galaxies of various types. The dotted line indicates the star forming main sequence from $z\sim0$ \citep{2007A&A...468...33E}. Super spirals, lenticulars and post-mergers are taken from \citet{Ogle_2019}. \citet{2017ApJS..233...20L} indicates the isolated massive spiral galaxies. Galaxies with different quenching stages like the star forming (SF), centrally quenched (cQ), mixed (MX), nearly retired (nR), fully retired (fR) and quiescent-nuclear-ring (QnR) taken from \citet[][]{2021A&A...648A..64K} have also been shown. UGC~12591 with $\rm M_{\star} = 1.6\times10^{11} \Msun$ and  $\rm SFR = 0.105 \Msun yr^{-1}$ from {\fontfamily{qcr}\selectfont CIGALE}, falls  in the      region of highly quenched fR galaxies. For comparison,  another extremely massive and fast rotating spiral J2345-0449 from \citep[][]{2014ApJ...788..174B,2021A&A...654A...8N} and  other massive spirals like NGC~1961 \citep[][]{2016MNRAS.455..227A} and NGC~6753 \citep[][]{2013ApJ...772...97B} are also plotted in both the graphs. {Right:} Here  the specific star formation rate (sSFR in ${\rm yr^{-1}}$) versus the stellar mass ($\rm M_{\star}$ in \Msun) is  shown for star forming ultraviolet luminous galaxies taken from \citet{2007ApJS..173..441H}.  On both panels data point labeled CIGALE  and MAGPHYS represent the estimates for UGC~12591 with sSFR and stellar mass from {\fontfamily{qcr}\selectfont CIGALE} and {\fontfamily{qcr}\selectfont MAGPHYS} respectively. Roughly $\rm log_{10}~sSFR > -10.5$ is the region of galaxies with active star formation, $\rm -11.5 < log_{10}~sSFR < -10.5   $ corresponds to green valley and  galaxies with  $\rm log_{10}~sSFR < -11.5$ represents the region of truly quiescent galaxies \citep{2016ApJS..227....2S}. It can be seen that galaxy UGC~12591 occupies a position in the lower right due to its highly quenched state where  another  massive and  quenched spiral  galaxy J2345-0449 (with AGN jets) is also located \citep[][]{2014ApJ...788..174B,2021A&A...654A...8N}. 
     NGC~1961  and NGC~6753 are located at  the lower end of star forming galaxies' main sequence.      }
     \label{fig:sfr_mstar}
\end{figure*}

\subsubsection{SED derived star formation rate}

We calculated the integrated star formation rate of UGC~12591 by SED fittings and obtained 
$\rm SFR_{SED} = 0.105\pm0.007  \Msun yr^{-1}$ from {\fontfamily{qcr}\selectfont CIGALE} and $\rm SFR_{SED} = 0.210^{+0.037}_{-0.097} \Msun yr^{-1}$ from {\fontfamily{qcr}\selectfont MAGPHYS} (Table~\ref{tab: magphys_tab}).  \citet{2017ApJS..233...20L} estimated the SFR of  $1.17\pm 0.13 \Msun yr^{-1}$ with WISE $22\micron$ data. Our estimate is significantly lower  and probes  recent star formation as it takes into account  the UV, NIR and  nebular fluxes. The star formation history (SFH) of UGC~12591 over the past $\rm 10^{7} - 10^{8}$ yrs  is extremely  low and does not differ much from the present day instantaneous star formation rate (Table~\ref{tab:cig_tab}). Evidently, all major growth in stellar mass should have happened before this time period and for at least past  $\rm 10^{8}$ yrs the galaxy has remained in the quenched state as a sterile,  `red and dead' galaxy as is evident in Fig.~\ref{fig:sfr_mstar}.

\subsubsection{FUV derived star formation rate}

The FUV emission is a good tracer of young stellar population and SFR upto $\sim 100$ Myr. We estimate the SFR using the FUV magnitude 20.18$\pm$0.25 mag (AB) from the GALEX catalog. The magnitude is corrected for dust extinction assuming $\rm R_{FUV}$ = 8.01 \citep{Walletal2019} and E(B-V) = 0.095 in the direction of UGC 12591 obtained from IPAC dust maps. The corrected FUV magnitude of UGC~12591 is 19.42$\pm$0.25 mag (AB) corresponding to $\rm F_{\nu}$ = $(6.19\pm1.42)\times10^{-28}~\rm ergs~s^{-1}~cm^{-2}~Hz^{-1}$ and luminosity $\rm L_{FUV}=(7.04\pm1.61)\times10^{26}~{\rm ergs~s^{-1}~Hz^{-1}}$. We then use the \citet{Kennicutt98} relation to estimate FUV SFR;

\begin{equation}
    \rm SFR_{FUV}\,(\Msun yr^{-1}) = 1.4 \times 10^{-28} L_{FUV}\, (ergs\,s^{-1}\,Hz^{-1})
\end{equation}

The estimated SFR is 0.098$\pm$0.022 $\rm M_{\odot}yr^{-1}$ showing exceptionally low  recent star formation activity, consistent with SED derived SFR above. 

\subsubsection{MIR derived star formation rates}

Here, we use the WISE $\rm 12 \micron$ (W3) and WISE $\rm 22 \micron$ (W4) fluxes \citep[][]{2013wise.rept....1C} (Table~\ref{tab:flux_data_ned}) in order to estimate the star formation rate. We use the relations provided by \citet[][]{2013AJ....145....6J},

\begin{equation}
   \rm{ SFR_{W3} (\Msun yr^{-1}) = 4.91\times10^{-10} \nu L_{12}(\Lsun)}\\
\end{equation}
\begin{equation}
   \rm{ SFR_{W4} (\Msun yr^{-1}) = 7.50\times10^{-10} \nu L_{22}(\Lsun)}
\end{equation}

These relations give us the estimated SFR in W3 as $\rm 0.638\pm0.051 \Msun yr^{-1}$ and in W4 as $\rm 0.345\pm0.043 \Msun yr^{-1}$. $\rm \nu L_{12}(\Lsun)$ and $\rm \nu L_{22}(\Lsun)$ are 12 and 22 $\rm \micron$ fluxes
respectively in \Lsun units.

\smallskip
Table~\ref{SFRS} shows a compilation of estimated SFRs by different 
methods. Each method has   it's  associated  systematic and measurement
uncertainties, so  different methods  may not  strictly  provide the
same result.  However these estimates  provide a robust
set of constraints about the most likely  range of the integrated,  
on-going and  recent past star formation rates in UGC~12591.  These values range
within $\rm SFR = 0.1  - 0.6 \, \Msun yr^{-1}$, consistent with a  very small star formation rate 
for this extremely massive galaxy.  The star formation rates are  all lower than that estimated by \citet[][]{2017ApJS..233...20L} based on a single MIR color. 
More observation of the target is needed in $\rm \gtsim 100 \micron$ wavelength in order to put better constraint on the star formation rate and the SED fitting in the region of FIR and beyond.

\begin{table}
    \centering
    \caption{A compilation of the estimated star formation rates by different methods.}
    \label{tab:all_sfr}
    \begin{tabular}{lc}
    \Xhline{2.7\arrayrulewidth}
    Method & Star Formation Rate  ($\Msun yr^{-1}$)  \\
    \hline
    SED Fitting ({\fontfamily{qcr}\selectfont CIGALE}) &  $0.105\pm0.007$ \\ SED Fitting ({\fontfamily{qcr}\selectfont MAGPHYS}) & $0.210^{+0.037}_{-0.097}$\\
    GALEX FUV & $0.098\pm0.022$\\
    WISE  W3  (12 micron)& $0.638\pm0.051$\\
    WISE W4 (22 micron) & $0.345\pm0.043$ \\
    \hline
    \end{tabular}
    \label{SFRS}
\end{table}

\subsubsection{Why is the SFR so low ?}

In Fig.~\ref{fig:sfr_mstar} it can be seen that UGC 12591 when compared to the main sequence of star forming galaxies \citep{2007A&A...468...33E} falls  at least 2 dex  below
this line due to its very low SFR.  The optically luminous massive spirals and lenticular galaxies from \citet{Ogle_2019} with $M_\star \ga 10^{11}\Msun$ also fall below the main sequence  line,  but are placed much above UGC 12591. Some of the massive, isolated spiral galaxies from \citet{2017ApJS..233...20L} with $M_\star \ga 10^{11}\Msun$ have also been shown. These isolated galaxies fall in the lower right region of the plot. It is interesting to note that UGC~12591 falls farther in the lower part of the diagram, in the region predominantly occupied by galaxies with nearly or fully quenched  star formation activity of \citet[][]{2021A&A...648A..64K}.  

On the right hand side of Fig.~\ref{fig:sfr_mstar} it can be seen that when compared to sequence of UV bright star forming galaxies, UGC 12591  is located in the lower-most region of  quiescent galaxies showing the lowest specific 
sSFR of all. In the same region is located J2345-0449, another very massive and highly rotating spiral galaxy studied by \citet[][]{2014ApJ...788..174B, 2021A&A...654A...8N}. The position of both  galaxies is in stark contrast with the ultraviolet luminous galaxies  of \citet{2007ApJS..173..441H} which show far more recent star forming activity for the same mass range. The low SFR
of both these very massive galaxies ($\rm M_{\star} >  10^{11}\Msun$)
located in sparse environment  show  that galaxy's mass and local environment is responsible to quench star formation. We plot  two other massive spirals  NGC~1961 and NGC~6753 which have low, partially quenched  sSFR, in accordance with their large mass  $\rm  log(M_{\star}) > 11.5$.
There are still very few studies of  cold molecular gas content of galaxies falling    below the main sequence of star formation, and a general
perception is that these systems are devoid of gas. However a 
detailed study of their molecular gas content is a key element for understanding the reason behind their remarkably low SFR.

UGC~12591 is one of the rare, most peculiar galaxies in the Universe having very high rotational velocities at around $500 \kms$ \citep{1986ApJ...301L...7G}, almost twice of the rotational velocity of the Milky Way and one of the fastest measured to date. UGC~12591 is isolated and the nearest satellite galaxy to it is 1.09 Mpc away. \citet{2004MNRAS.354..851R} discuss possible scenarios of formation of the early-type isolated galaxies like early epoch collapse, merging event, accretion etc. The dust lanes of UGC~12591 indicate a past merging or  accretion event.

Any merging event has drastic effect on the shape and size of the parent galaxy as well as its star formation. In most cases  the star formation is triggered due to collision \citep{1988ApJ...325...74S} and takes place in the nuclear region of $\rm 10^{2} - 10^{3}~pc$ of the galaxy. If there is no hot halo then the galaxy becomes bulge dominated after merger. Otherwise, hot halo helps inflow of the cold gas which again helps in the regrowth of the disc by increasing SFR as shown by \citet{2015MNRAS.452.4347K}. In contrast to that \citet{2011MNRAS.415.3750M} did a study on the star forming efficiency during mergers and find that the presence of a hot halo reduces the efficiency of star formation as compared to absence of a hot halo during the merging event. \citet{2012ApJ...755..107D} using XMM-Newton  detected a hot gaseous halo in UGC~12591. So, it can be said that there was a time when the galaxy might  have gone through higher star formation rate. But from SED fitting it is inferred that the total mass of the young stars is negligibly small. That indicates the recent SFR ($\rm t_{age} < 10^{8}$ years) is extremely low and the object most probably has crossed an era of violent quenching of its star formation.
How this  remarkable quenching of star formation could have happened  early in its evolutionary  history and  why   star formation did not resume at later stages are still  open issues which needs more  detailed study. Nevertheless, we discuss some possibilities;

The reason for the low star formation rate in early-type galaxies like UGC~12591 could also be related to  AGN activity of central black hole (further discussed below), or related to the structural properties.  \citet{2013MNRAS.432.1914M} finds that early-type galaxies form stars 2-5 times less efficiently than the spirals due to the reason that gas discs of early-types are more stable against star formation because of which they do not create high density clumps capable of star formation. \citet{2005A&A...444L...9M} show
that radio-loud phase of AGN activity can hamper the star formation by heating up the interstellar gas making it metal rich. \citet{2014MNRAS.438.1870D} describes two mechanisms for the suppression of star formation in massive galaxies; internal quenching by a compact bulge, specifically at high red shifts, and quenching due to hot halos at low red-shift.

\subsection{Central black hole mass and  AGN feedback}
\label{subsec:bhm}

Here we investigate the possible quenching of star formation due to growth of
a supermassive black hole. The mass and accretion state of this black hole
are key factors which decide its radiative or kinetic influence on the surrounding medium. 
We  estimated the   nuclear black hole mass from the tight ${\rm M_{bh}- \sigma_{c}}$ relation for 
a sample of E, S0 and S types of galaxies as given in \citet{2009ApJ...698..198G}, following the equation,

\begin{equation}
\log(\frac{\rm M_{bh}}\Msun) = \alpha + \beta  \log(\frac{\sigma_{c}}{200\, {\rm km s^{-1}}})
\end{equation}

This relation is calibrated using dynamically detected central black hole
masses and has an intrinsic rms scatter of $\rm \epsilon = 0.44 \pm 0.06$. 
Here $\alpha=8.12\pm0.08$ and $ \beta=4.24\pm0.41$ and $\sigma_{c}$ is the central velocity dispersion, taken as $288.0\pm23.9$ {$\rm km~s^{-1}$}
from HyperLEDA \citep{2014A&A...570A..13M} \footnote{http://leda.univ-lyon1.fr/}. The black hole mass is obtained as $\rm M_{bh} = (6.18\pm2.61)\times 10^{8} \Msun$. 
Similar  mass   is obtained ($\rm M_{bh} = (8 \pm 1)\times 10^{8} \Msun$) if we use the velocity dispersion $306.0\pm7.0$ {$\rm km~s^{-1}$} quoted in \citet[][]{2015ApJS..218...10V}.
UGC~12591 being a  highly bulge dominated galaxy, the  ${\rm M_{bh} - \sigma}$ relation should give a good estimate of the black hole mass \citep{2013ARA&A..51..511K}. 

\citet{2019ApJ...876..155S} analysed a sample of 84 early-type galaxies with directly measured SMBH masses and show how black hole mass is  tightly scaled  to spheroidal as well as total stellar masses. 
For UGC~12591 if we use   $\rm M_{bh} = 6.18\times10^{8} \Msun$ 
then we obtain its spheroid/bulge stellar mass $\rm M_{\star,sph} = 4.4\times10^{10}\Msun$ and total galaxy stellar mass $\rm M_{\star,gal} = 1.1\times10^{11} \Msun$ (\citet[][Eq.~12,14]{2019ApJ...876..155S}). 
This estimate of  $\rm M_{\star,gal}$  brackets the  stellar mass calculated from SED fitting in the range $9.1\times10^{10} - 1.6\times10^{11}\Msun$  and  $\rm M_{\star,sph}$ mass is close to  our  estimate of V band bulge mass (Table~\ref{tab:cig_tab}, \ref{tab: magphys_tab} and Table~\ref{tab:luminosity_mass}). This shows that UGC~12591 hosts a massive black hole of  $\rm \approx 1.4 \%$ of bulge mass while 
$\rm \approx 28-46 \% $ of the total stellar mass  resides in the bulge itself, consistent with the highly bulge dominated nature of UGC~12591. 

The above  black hole mass  ${\rm M_{bh}}$ is in the range of super massive black holes,  reasonable enough for a  massive bulged galaxy like UGC~12591. The  origin and growth of a black hole within a galaxy this  massive remains an unsolved problem \citep{2013ARA&A..51..511K}. Even though the sphere of influence of a black hole is small when compared to the size of a galaxy, a SMBH can effectively regulate the star formation within the galaxy. It has been found that more massive is the bulge more massive is the black hole  and the host galaxy will be redder in colour. In addition, the black hole may  have a powerful effect on the evolution of a galaxy, although there is no good observational corroboration of the fact. However, recently \citet{2018Natur.553..307M} from a study of 74 massive galaxies obtained  a striking correlation between the black hole mass and star formation rate of the galaxy. 
The galaxies with over massive black holes have SFR which is quenched much earlier than the SFR of the galaxies with  under massive black holes. At the same time, the galaxies with over massive black holes had  more star formation going on at a faster rate in the past than the other category of galaxies. They conclude that the higher baryon cooling efficiency at high redshift would play a major role in growth  of supermassive black holes, feeding the primordial seeds of the black holes with gas and forming more stars, in agreement with  quasar observations at higher
redshifts. The stellar mass formed at  $\rm {\it z} \approx 5$ in over-massive black-hole galaxies is about $ 30\%$ more than those in under-massive black-hole galaxies. 

Finding clues from its present properties, we propose that the progenitor of UGC~12591 may have gone through an intense starburst phase in the past. The subsequent suppression of the star formation in the galaxy was possibly mediated by the  intense   AGN activity of the  growing  SMBH. From SED fitting we found  that this
SMBH at the present moment is quiescent, i.e. neither we see large  ($\rm > 1 kpc$) radio jets nor is the  black hole contributing significantly to the  mid-IR flux,  ruling out the presence of strong  radiative feedback from a  bright AGN.  This,  and absence of large radio jets  implies possible feedback of this AGN  is via  small-scale jets ($<1 $ kpc) or uncollimated winds/outflows    powered by  low Eddington rate, hot accretion flow on the SMBH \citep[][]{2014ARA&A..52..529Y}.

Since even in low frequency surveys (like the 150 MHz TGSS ADR; \citet[][]{2017A&A...598A..78I}) no trace
of a previously active, currently relic or fading radio lobes are seen, and furthermore no fresh  episode of recent  jet activity is visible, we assume that  the AGN has remained  quiescent for at a least a few $10^{8}$ years, exceeding the radiative life-time of typical radio galaxies \citep[][]{2020A&A...638A..34J}.
However, there is a strong possibility that  in distant past this galaxy may have passed through a phase in which the  accretion rate of matter around the black hole was high, resulting in far stronger AGN activity giving rise to energetic outflows (in quasar mode) or  powerful relativistic  jets  (in radio mode) which made the  environment of the galaxy  hostile of forming  new stars. Numerical simulation results of black hole
growth via mergers in early Universe match this scenario \citep[][]{2005Natur.433..604D}. 
Recent CO observations of a
massive, Mpc-jet  launching  spiral J2345-0449 showing strong SFR quenching \citep[][]{2021A&A...654A...8N} and  state-of-the-art relativistic hydrodynamic
simulations of  jets  propagating in a dense ISM
\citep[][]{2016MNRAS.461..967M,2021MNRAS.508.4738M} provide good
support for this scenario, while future radio and X-ray telescopes have great  potential
of revolutionizing this field \citep{2018ApJ...859...23N}.

Other than AGN activity there are other methods discussed in literature by which the star formation can be quenched. \citet{2018A&A...609A..60K} finds that the presence of a stellar bar can effectively reduce the star formation by a factor of 10 in less than 1~Gyr time. \citet{2009ApJ...707..250M} suggest `morphological quenching',  where in early-type galaxies the quenching of star formation can result from growth of a stellar spheroid/bulge by mergers. During the transition of stellar disc to spheroidal shape,  strong shear boosted by the steep gravitational potential gradients of their bulges may  lower SFR in such galaxies.




\begin{figure*}
\centering
\includegraphics[scale = 0.35]{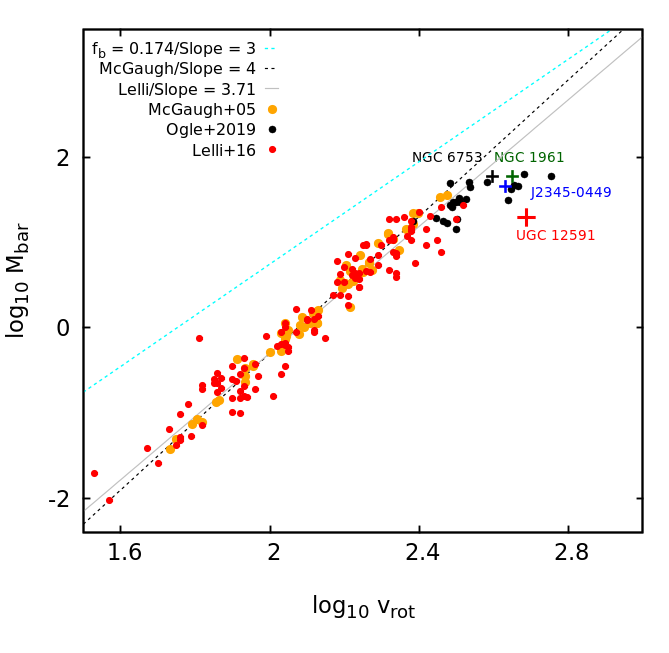}
\includegraphics[scale = 0.35]{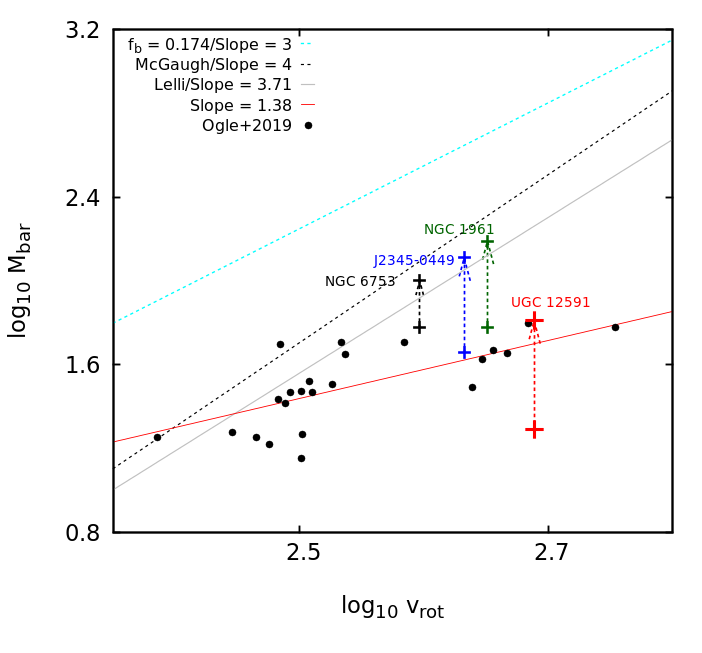}
\caption{In these  plots the  baryonic mass and the rotational velocity of  UGC~12591 is compared to  some     other galaxies. {Left}: A baryonic Tully-Fisher (BTF) relation is shown,  where ${\rm M_{bar}}$ is the  baryonic mass  in  units of $10^{10} \Msun$ and ${\rm v_{rot}}$ is the maximum (flat) rotational velocity in ${\rm km\, s^{-1}}$ in both the plots. Galaxy data from \citet{2005ApJ...632..859M} and \citet[][]{2016AJ....152..157L} are plotted with the linear fit in  lower rotation velocity/mass  regime  $\rm \approx v_{rot} \ltsim 300 {\rm km\, s^{-1}}$ deriving similar slopes of $ 4.00$ and $ 3.71$ respectively (dashed and solid lines). The data for highly  rotating, massive  super-spirals from \citet{2019ApJ...884L..11O} have also been plotted (black points), where most of the galaxies have rotational velocities $ > 300 {\rm km \, s^{-1}}$. The rotational velocity of UGC 12591 is taken as 488.4 ${\rm km\, s^{-1}}$ from Hyperleda \citep{2014A&A...570A..13M}. We find that the tight, linear BTF relation starts  to turn over (flatten) for galaxies with highest $\rm v_{rot}$, implying a significant baryonic shortfall in massive halos,  which is also mentioned in \citet{2012ApJ...755..107D}, \citet{2019ApJ...884L..11O} and   \citet{2015MNRAS.448.1767C}.
In these plots the radio-loud (hosting Mpc scale radio jets) spiral  galaxy  J2345-0448 is  a rare  exception  where significant     mass of the cosmic  baryons are found (see Fig.~\ref{fig:fb_vrot}) in an luminous, extended  hot X-ray halo \citep[][]{2014ApJ...788..174B,2021MNRAS.500.2503M,2015MNRAS.449.3527W}. {Right:} Here we fit the super-spiral sample from \citet[][]{2019ApJ...884L..11O} along with UGC~12591 and other massive spiral galaxies J2345-0449, NGC~1961 \citep[][]{2016MNRAS.455..227A} and NGC~6753 \citep[][]{2013ApJ...772...97B} to find out how much their BTFR slope is differing from that  in the linear regime of lower mass galaxies in left figure. The slope is $\rm 1.38\pm0.34$,  which is much flatter compared to  $\rm \sim 3.7 - 4$. For the four most massive galaxies mentioned,  the upper data points are total baryonic mass, including the hot gas component and the lower points represent  the  'condensed' baryonic mass, excluding the hot gas budget. The baryonic masses of hot gas components are estimated within the virial radius of the galaxies from published X-ray observations. It can be seen that after inclusion of hot X-ray halo baryons in J2345-0449, NGC~1961 and NGC~6753,  they all fall exactly  on  the  BTF relation   of slope $\rm \sim 3.7 - 4$. On the contrary, UGC~12591 still falls much  below of that line even after including  the hot halo gas mass extrapolated upto its virial radius. In both the plots we have shown the $ \rm M_{bar} \propto  v_{rot}^{3}$ power law (the light-blue dotted line) for cosmic baryon fraction $\rm f_{b} = 0.174$ following \citet[][]{McGaugh_2012,2016ApJ...816L..14L}.}
\label{fig:btfr_fit}
\end{figure*}

\subsection{Baryon census in relation to the galaxy halo mass and SFR}

\subsubsection{The Baryonic Tully-Fisher relation} \label{subsec:btfr}

The Baryonic Tully-Fisher relation (BTFR; \citet{2000ApJ...533L..99M,2005ApJ...632..859M}) is a remarkably tight  power-law correlation connecting the baryonic mass and the 
circular rotational velocity (halo mass and angular momentum) 
of spiral galaxies which is  
 measured to follow a single power-law over a very wide
range in galaxy mass, e.g. 
\citep[][]{2001ApJ...550..212B,2016ApJ...816L..14L,2001ApJ...563..694V,2016A&A...593A..39P,2006ApJ...653..240G,2021MNRAS.508.1195P}. Currently, the origin of BTFR is not fully understood, but it holds the key to   understanding  galaxies as
physical objects, their dark-matter, baryon and stellar content, and most
importantly,  how they formed. 
\vskip 0.15cm
The BTFR can be expressed  in the following way,
\begin{equation}
{\rm M_{bar}} = \gamma {\rm v_{rot}}^{\beta}
\end{equation}
Where, ${\rm  M_{bar}} $ is the total baryonic mass of the galaxy, which includes the mass of the stars, ${\rm M_{\star}}$ and mass of the gases, ${\rm M_{gas}}$ (molecular, atomic  and
ionized). Here $\gamma$ is the normalising parameter 
and $\beta$ is the slope in the linearised equation using logarithmic scale. 

SED fitting  for UGC~12591 has  yielded the stellar mass, ${\rm M_{\star}} = (1.6\pm0.1)\times10^{11} \Msun$ and the  gas mass $\rm M_{gas} = (7.1\pm0.5) \times 10^{10} \Msun$ (Table~\ref{tab:cig_tab}). As UGC~12591 does not have CO observations to pin down the cold molecular gas mass, nor optical spectroscopy for the ionized gas content, we do not consider the gas mass provided by {\fontfamily{qcr}\selectfont CIGALE}, but estimate it via method outlined below in our baryon budget.

In order to estimate the atomic HI gas mass of the galaxy from single dish line
profile,  the following equation has been used  (applicable for
optically thin condition),

\begin{equation}
{\rm M_{HI}} = 2.356\times10^{5} \times D_{L}^{2} \times (1 + z)^{{-2}} \times \int F dv
\end{equation}

Here, ${\rm M_{HI}}$ is the HI mass in  \Msun, $\rm D_{L}$ is the luminosity distance in Mpc and $\int F dv$ is the integrated  HI line flux in the unit of Jy-km ${\rm s^{-1}}$. Using observed  $\int F dv$ = 2.45  Jy km ${\rm s^{-1}}$ \citep[][]{1986ApJ...301L...7G} and  $\rm D_{L}$ as 97.5 Mpc, the total HI gas mass ($\rm M_{HI}$) is estimated at  $5.2 (\pm 1.6) \times10^{9} \Msun$. 
The uncertainty in HI mass is obtained by the method suggested by \citep[][]{2006MNRAS.372..977D}. We find that the HI mass is high but in the  region of  most massive galaxies,  with $\rm M_{HI}/M_{\star} \sim 0.04 $, and  lies  beyond the  break   at $\rm M_{\star} \sim 10^{9} \Msun$ on the
$\rm M_{HI}$ - $\rm M_{\star}$ plane \citep[][]{2015MNRAS.447.1610M}

Now in order to also include the molecular gas content we use the relation provided by \citet[][]{1989ApJ...347L..55Y} which shows the ratio of molecular to atomic gas mass of a S0/Sa hybrid galaxy to be $\rm 4.0\pm 1.9$. This gives us the molecular gas mass $\rm M_{H_{2}}$ of UGC~12591 as $\rm (2.1\pm1.2)\times10^{10} \Msun$. Adding  contribution of Helium and metals into account, we obtain the total gas mass $\rm M_{gas}$ as ${\rm M_{H_{2}} + M_{He}} + {\rm M_{HI}} = 1.38\times {\rm (M_{H_{2}} + M_{HI}})$ \citep[][]{2014A&A...563A..31R}, yielding $\rm M_{gas} = (3.6\pm1.7) \times10^{10} \Msun$. The contribution of ionized gas mass to the total gas mass is neglected, 
because very small star formation rate and lack of  strong ionizing UV radiation of hot, young stars 
(short ward of 912  \AA) or shocks would render ionized gas fraction negligible compared to the atomic and molecular gas mass. 
Future observations of recombination lines of Hydrogen should provide much better constrains on it.



Summing up all this disc baryon census ($\rm M_{gas} + M_{\star} + M_{dust}$) with the  X-ray halo hot gas mass ${\rm M_{gas,hot}}$ of $4.5\times10^{11} \Msun$ within 500 kpc from the centre of the galaxy \citep{2012ApJ...755..107D}, the total baryonic mass of the galaxy becomes $(6.46\pm0.20)\times10^{11} \Msun$. In Fig.~\ref{fig:btfr_fit}, we have plotted both the total baryonic mass with hot gas contribution 
and  without  hot gas on the  BTFR  along with  different samples of galaxies mentioned in figure caption.  Two  best fitting lines  of  the form   $\rm M_{bar} \propto v^{\beta}_{rot}$ are also shown  as  determined by  \citep[][]{2005ApJ...632..859M} ($\rm \beta = 4.0$) and \citep[][]{2016AJ....152..157L} ($\rm \beta = 3.71$). We further discuss the  implications of   BTFR below.

\subsubsection{Where are most of the baryons in super massive ($\rm > L_{\star}$) galaxies? }

Generally the baryons census  around massive $\rm > L_{\star}$ local galaxies reveal  much less observable baryons than that is expected from Big Bang nucleo-synthesis and Cosmic Microwave Background constraints (see
\citet{2017ARA&A..55..389T} for a review). Previous studies have shown (albeit for  a handful of galaxies so far) that only a  small fraction  ($\rm \sim 10-30\%$) of these `missing baryons' are located in the warm-hot ($\rm \sim 10^{6}$ K) circum-galactic coronal gas  around the massive galaxies,  possibly extending upto the virial radius of their dark matter halos \citep[][]{2012ApJ...755..107D,2021MNRAS.500.2503M,
2015MNRAS.449.3527W,2011ApJ...737...22A,2016MNRAS.455..227A,2018ApJ...862....3B,2013ApJ...772...97B}. 
UGC~12591 is an excellent target to probe the `missing baryon' problem because it is unusually massive and situated in a low density environment which causes lower interaction with the intra-cluster medium and other galaxies. 

The position of most massive spiral galaxies on the Baryonic Tully-Fisher  relation  should  provide some fundamental insights
into the missing baryon problem  and  galaxy formation. 
The  remarkably tight  baryonic Tully-Fisher relation  \citep[][]{2016ApJ...816L..14L} shown in Fig.~\ref{fig:btfr_fit}
of slope $3.7-4.0$ connecting  ${\rm M_{bar}}$, the total baryonic mass of the galaxies and the halo mass, represented by ${\rm v_{rot}}$ the maximum rotational velocity,  implies that on an average a nearly constant fraction ($\sim0.4$) of all baryons expected in a halo are ``condensed'' onto the disc  of  rotationally supported galaxies \citep[][]{2014AJ....147..134Z}. The dashed blue line shows the $\Lambda$CDM prediction assuming that galaxies contain all available cosmological baryons.
A constant slope ($\beta \sim 4$) and  remarkable tightness of BTFR suggest a self-similar  scaling mechanism
for  rotating discs operating across a wide range of halo masses. However,
it has been suspected \citep[][]{2012ApJ...755..107D,2019ApJ...884L..11O,2015MNRAS.448.1767C}  that for  highest mass $\rm L > L_{\star}$ galaxies
($\rm M_{\star} \gtsim  10^{11} M_{\sun}$) the linear BTF relation starts  to turn over/flatten  at  highest $\rm v_{rot}$, implying a significant baryonic shortfall of  massive galactic discs.  This part of the BTFR has remained the least studied so far. We reexamine and  confirm this trend  as shown in  Fig.~\ref{fig:btfr_fit} by including UGC~12591 and some of  the most massive disc galaxies known in the literature. For
these galaxies the derived BTFR slope is $\rm 1.38\pm0.34$,  which is much flatter compared to  $\rm \sim 3.7 - 4$, for lower mass halos. 

Furthermore, we find that when the expected  baryonic mass from   a  linear fitted   BTFR of slope $\rm \sim 4$ is assumed  to be present in UGC~12591, and  assume
 $ \rm M_{200,xray} = 1.9\times10^{13} \Msun$ \citep{2012ApJ...755..107D},  we can  restore the galaxy's  baryonic mass fraction up to $ f_{b} = 0.147$,  near to the cosmological closure value  0.174.
In that case  about four  times more baryonic  mass possibly resides in an  extended galactic halo  over  what has been observed in  various other baryon components so far, including those in the X-ray halo (Fig.~\ref{fig:btfr_fit} right panel).
Thus, to account for these  so called `missing baryons'  this would imply a baryon shortfall of  $\sim$ 77\% in this galaxy. We show in Fig.~\ref{fig:btfr_fit} that a similar baryon shortfall with respect to the linear BTFR is also indicated  for NGC~1961, NGC~6753, J2345-0449 and for several other highly rotating,  massive  super-spirals taken from \citet{2019ApJ...884L..11O}.  We show that after inclusion of hot X-ray halo baryons observed in J2345-0449, NGC~1961 and NGC~6753,  they all fall exactly  on  the  BTF relation   of slope $\rm \sim 3.7 - 4$. On the contrary, UGC~12591 still falls much  below of that line even after including  the hot halo gas mass extrapolated upto its virial radius. Many phenomenon can drive the baryon loss of galaxies such as supernova blowout, AGN feedback via a quasar mode wind or a via powerful radio jets, as reviewed in \citet[][]{2017ARA&A..55..389T}.

Why do the  most massive  galaxies  may deviate from the tight, linear  fitting BTFR law? 
One interesting possibility why the linear BTFR law breaks/flattens for most massive halos is because  a  comparatively  lesser  fraction of
halo baryons in these galaxies are able to condense into discs  and form  new stars or form  massive  neutral hydrogen discs, as
we find in this work and further discussed below (Section~\ref{Subsec: massive_spg}). This could indicate a  deviation or
perturbation in the  self-similar scaling relation due to non-gravitational processes such as AGN feedback, supernova blowout, bulge formation  or  a longer radiative cooling time of
baryons in CGM. Interestingly, \citet[][]{2019A&A...622A..17S} showed that the central SMBH in most massive
galaxies ($\rm M_{\star} > 10^{11} \Msun$) are always switched on at various levels of AGN activity in radio wavelengths, in line with models in which these  AGN are essential for maintaining the quenched state of galaxies at the centres of hot gas halos. The role of specific angular
momentum and  stellar bulge of the  disc in deciding  this deviation at the highest galactic rotation
speeds  also needs to be 
understood.
It remains to be seen if this departure from linear BTFR is a generic feature for most massive galaxies, using more data on these rare giants. This is a powerful clue, as the exact physical process inhibiting star formation in massive  rotating discs is still unclear and  this  signal presents an outstanding  challenge  for future investigations.  

The  rotation curve for UGC~12591 is constrained to be flat only within about  40 kpc from the nucleus \citep[][]{1986ApJ...301L...7G}. 
Rotation curves are the major
tools for determining the mass distribution in spiral galaxies,
e.g. \citep[][]{2001ARA&A..39..137S}. 
If a flat rotation curve and  Kepler's law  is assumed to hold true upto  80 kpc from the centre  (the maximum extent of soft X-ray halo), then one can obtain an interior  gravitational mass ${\rm M_{g,tot}} = 4.44\times10^{12} \Msun$, i.e. ${\rm M_{xray-halo}} \sim 0.1 \, {\rm M_{g,tot}}$ within 80 kpc. Accounting for the  baryonic mass of the
hot X-ray halo ${\rm M_{gas,hot}}$ of $4.5\times10^{11} \Msun$ and the total baryonic mass in disc found from our SED-fitting,  one can estimate the  baryonic fraction to be $\rm 
f_{b} = 0.145$  which is close to the cosmological mean value of 0.174 \citep[][]{2007ApJS..170..377S}. Thus, within the X-ray halo radius,
the baryon fraction is near closure and a significant fraction of these baryons reside in a hot gaseous phase, rather than forming stars.

One should treat this result with caution as the HI rotation curve of UGC~12591 has not been measured upto  80 kpc from the
center and we have no information about the distribution of dark matter mass $\rm M_{g,tot}$. Future spatially resolved, deep 21cm radio observations may measure the
rotation curve more accurately, probing the gravitational potential,  as well as it may reveal the anomalous HI gas structures (rings, outflows, warps)  with high angular resolution  at the outskirts of this galaxy.

What is the total mass budget within $\rm R_{200}$? From a tight correlation between black hole mass and the total gravitational mass found by \citep{2009ApJ...704.1135B} the  gravitational mass within virial radius ($R_{200}\approx 550$ kpc) $\rm M_{200}$  can be calculated  from the following,  

\begin{equation}
\log{\rm M_{bh}} = (8.18\pm0.11) + (1.55\pm0.31)[-13 + \log{(\rm M_{200}/M_{\odot})}]
\end{equation}

To use this correlation we need the central black hole mass.
Now, using the ${\rm M_{bh}}$ obtained from the  ${\rm M-\sigma}$ relation and its
scatter,  we obtain the total halo mass
${\rm M_{200}} = (2.47\pm0.89)\times10^{13} \Msun$. This mass is close to the value calculated by \citet{2012ApJ...755..107D} of $ \rm M_{200,xray} = 1.9\times10^{13} \Msun$  from the X-ray halo data assuming  hydro-static equilibrium condition. If one uses this value of $\rm M_{200}$ and use observed $\rm M_{bar}$, then the estimate of 
mass fraction  in baryons  is only $\sim 0.026$ out to ${\rm R_{200}}$,  as against the
cosmological value $\sim 0.174$, which is  extremely puzzling. The baryon mass budget reads; total baryon mass $4.3\times10^{12}\Msun$, observed baryon mass $6.46 \times10^{11}\Msun$, and yet-to-be observed
baryon mass $3.65\times10^{12}\Msun$. This  $\sim 85\%$ `missing baryon'  mass  (or possibly  the missing light) discrepancy creates a tension  with the standard  models of galaxy formation
\citep[][]{1978MNRAS.183..341W,1991ApJ...379...52W,2006ApJ...639..590F,2006ApJ...644L...1S} which demand a hot X-ray corona containing a large fraction  (at least $50 \%$) of cosmic baryons, thus
bringing us face-to-face to the most important question: where do most of the undetected cosmic baryons in UGC~12591 reside and in what form?    

There are several possibilities;
(a) The hot  corona  of  UGC~12591  is not
in hydrostatic equilibrium,  giving an erroneous estimate of  ${\rm M_{200}}$. This can not be ruled out as of yet and it would mean  that the halo gas is dynamically unstable, either expanding outwards or collapsing inwards.  However, If the 
X-ray halo gas is in  virial equilibrium, it will attain  the virial temperature of 
($\rm  T_{vir}$),  
corresponding to the  depth of gravitational potential well and  exhibit Keplerian rotational (virial) velocity $\rm v_{rot}$; 

\begin{equation}
\rm T_{vir}(eV) = \rm 34.47 \times (v_{rot}/100 \kms)^{2}
\end{equation}


For $\rm v_{rot} = 400 - 500 \kms$ (halo mass $\rm \sim 10^{13} M_{\sun}$) we find $\rm T_{vir} = 0.5 - 0.8 \, keV$, quite close to the
measured  X-ray value $\rm T_{xray} \approx 0.64\pm0.03  \, keV$ \citep[][]{2012ApJ...755..107D}. If not a  coincidence, the
X-ray halo gas is in viral equilibrium consistent with a  massive dark matter halo of total mass $\rm few \times10^{13}\Msun$ as suggested by  \citet{2012ApJ...755..107D}. This in turn implies the halo gas is possibly  shock heated after being accreted  from the cosmic-web, rather than recycled from the disc. The thermal and metal line cooling time of the quasi-hydrostatic X-ray halo gas is  large, in the range  $2.8 - 6.3$ Gyr for material within the cooling radius, and has the effective cooling rate of $\rm 0.15 - 0.21 M_{\sun}\, year^{-1}$ \citep[][]{2006ApJ...639..590F,2012ApJ...755..107D,2021MNRAS.502.2934K}.  The average estimated cooling time is much larger than the free-fall time, and therefore, it would require more than a Hubble time to build the entire stellar mass of galaxy by condensation of cooling gas from halo, thus disfavoring such a possibility. The dynamical state of the circum-galactic  gas would be revealed
in future sensitive spectroscopic observations from the detection of the `Warm Hot Intergalactic Medium (WHIM)' in the soft X-ray lines at $\sim 10^{6}$ K, and via the  detection of  even cooler gas through the UV/EUV lines of  metals \citep[][]{2021ExA....51.1043S,2021ExA....51.1013N}.

(b) a large repository of cosmic baryons is yet to be found  within ${\rm R_{200}}$ and
possibly beyond. What is the  state (density, temperature and metal content) of  these  baryons?  The X-ray corona of all the over massive spirals  detected so far 
show a  trend of declining temperature profile, and beyond virial radius their gas temperature would most likely be below $< 10^{6}$ K  where  EUV and FUV
wavelengths  provide the best diagnostics.
These baryons may exist  in a cold and low density  phase which makes it very difficult to observe them with current X-ray telescopes, hence these galaxies are prime targets for future  highly promising  exploratory  missions \citep[][]{2021ExA....51.1043S,2018ApJ...862....3B,2021ExA....51.1013N}.

Another possibility (c) is that actual ${\rm M_{200}}$ is far smaller than above, as we require only $\approx 4\times10^{12}\Msun$ to resolve the tension with
cosmological baryon fraction.  If true this would  imply that possibly
all the baryons are present within  $\sim100$ kpc; $\rm \sim4.5\times10^{11} \Msun$ ($\sim66 \%$) in
the X-ray halo and $\rm \sim2.4\times10^{11} \Msun$ ($\sim34 \%$) in stars and ISM,  and the rest is in
dark matter of mass $\rm \sim 3.3\times10^{12}\Msun$. 
In that case  the rotation curve may  show a steep decline in the halo region if  measurable at large galactocentric radii. Sensitive 21cm Hydrogen line observations can be an
excellent tool to probe this.
Since there are no robust observational estimates  of ${\rm M_{200}}$ as yet, one can not rule out this possibility. However in  the 
widely accepted standard cosmological paradigm we expect the  most massive galaxies to  form in most massive halos dominated by dark matter and
a massive dark matter halo has a strong dynamical and stabilizing influence on the disc. We again emphasize  that the total halo  mass of UGC~12591 is  the crucial parameter that is not well constrained, since it also depends on  extrapolation of the observed rotation curve  or X-ray gas profile to $\rm R_{200}$.


\subsubsection{Baryons in UGC 12591 and  Super-Spirals}

Super spirals are an interesting newly 
discovered galaxy population which are  extremely massive, fast
rotating  HI-rich disc galaxies with significantly high star forming rates \citep[][]{2019ApJ...884L..11O}. 
In terms of rotation velocity a few of them approach or even exceed the $\rm v_{rot}$ for UGC~12591. In Fig.~\ref{fig:sfr_mstar} and Fig.~\ref{fig:btfr_fit} we have included the super spiral sample of \citet{2019ApJ...884L..11O}. We 
find that compared to the super-spirals  which all lie close to
the main sequence of star formation, UGC~12591 has a strikingly low star formation rate. \citet{2019ApJ...884L..11O} note  that super-spirals show a break in the BTFR  at mass limit of $\rm > log(M_\star/\Msun) \sim 11.5 $ (see Fig.~\ref{fig:btfr_fit}). UGC~12591 confirms to the same trend with its $\rm log(M_\star/\Msun) \sim 11.2$ and  it falls in the region of highest
mass super spirals, well  beyond the  break in the linear BTFR,  which is quite evident from the figure. This indicates that the present SFR of a galaxy does not have a major effect on its position on the baryonic Tully-Fisher plane but it is the
total baryon mass which is most important. 


As discussed in \S~\ref{subsec:sfr} that in the past  UGC~12591 may have had a starburst phase which has been quenched exponentially over time. The main reason for a galaxy to stop forming stars is the unavailability of the necessary light elements like hydrogen and helium in its disc. So, the small star formation rate indicates the presence of more abundance of  metallic elements in its gaseous clouds which could have been thus enriched due to  past supernova explosions in the disc which in turn may have blown out baryons in the halo of the galaxy. In future a  sensitive spectroscopic measurement of  metal abundance  of the baryons in the circum galactic halos  will be necessary to test this model (see \citet[][]{2021ExA....51.1043S,2021ExA....51.1013N} and a review by \citet[][]{2017ARA&A..55..389T}).

\subsubsection{Condensed Baryon Fraction  and Baryons to Stars Conversion Efficiency}
\label{Subsec: massive_spg}

\begin{figure*}
    \centering
    \includegraphics[scale=0.35]{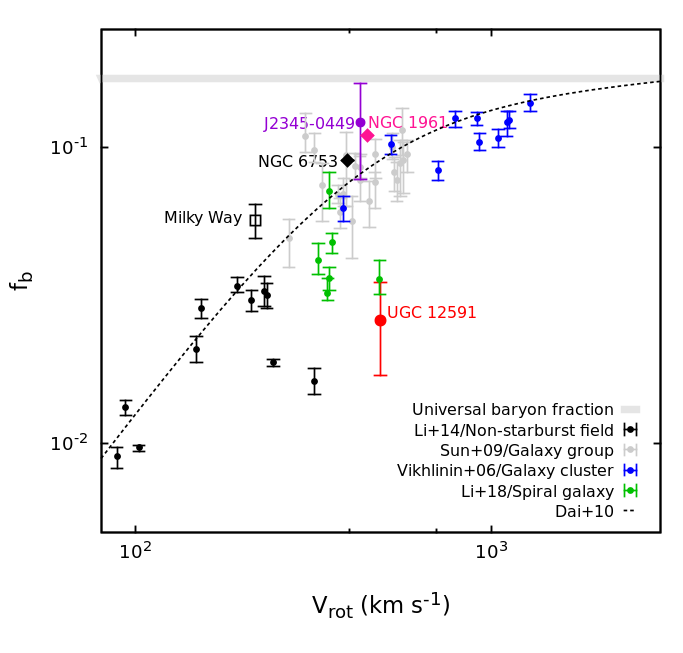}
    \includegraphics[scale=0.35]{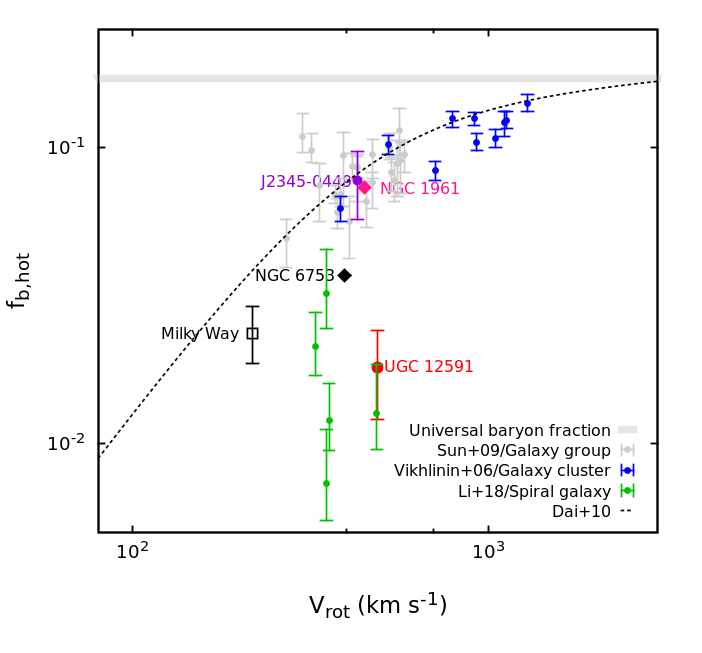}
    \caption{The baryon fractions of galaxies, groups and clusters with different rotational velocities have been shown along with UGC~12591. The left plot shows fraction of all baryons 
    $\rm f_{b}$ and the right plot shows the baryon fraction only in hot halo phase $\rm f_{b,hot}$ relative to the     cosmic baryon fraction (horizontal grey zone ~ $\rm 0.167-0.174$; \citet{2007ApJS..170..377S,2009ApJS..180..330K}). 
        The data points consist of non-starburst field spirals \citep[][]{2014MNRAS.440..859L}, spirals from \citet[][]{2018ApJ...855L..24L}, galaxy groups of \citet[][]{2009ApJ...693.1142S}, galaxy clusters \citep[][]{2006ApJ...640..691V} and the Milky way \citep[][]{2015ApJ...800...14M} with their uncertainties. The dotted line represents the power law model between $\rm f_{b}$ and $\rm v_{rot}$ presented by \citet[][]{2010ApJ...719..119D}. For comparison another extremely massive and fast rotating spiral galaxy J2345-0449 has been shown on  these graphs \citep[][]{2014ApJ...788..174B,2021MNRAS.500.2503M,2015MNRAS.449.3527W} that is seen to lie on the edge of the cosmic baryon fraction.  Massive spirals like NGC~1961  \citep[][]{2016MNRAS.455..227A} and NGC~6753 \citep[][]{2013ApJ...772...97B} can also be seen to be in the nearby region of cosmic baryon fraction. In contrast  UGC~12591 falls far below the  universal baryon fraction line, as it has baryon fraction out to $\rm R_{200}$ of only $\approx 0.026\pm0.009$, showing a baryon deficiency of $\sim 85\%$, signaling  a large tension with cosmological predictions.} 
\label{fig:fb_vrot}
\end{figure*}

For reasons still unclear, 
massive local galaxies like UGC 12591  have low specific star formation rate compared to other less massive spirals (see \citet{2018ARA&A..56..435W} for a review).  
However, interestingly,  \citet{2019A&A...626A..56P} in a study of nearby star-forming galaxies with $M_\star$ ranging from $10^{7} - 10^{11} \Msun$  found that some  massive spirals with $\rm M_\star \gtsim 10^{11} \Msun$ have turned most of the baryons in their halos into stars, without being subjected much to the quenching processes and there are hardy any missing baryons in them (see also \citet{2021arXiv211204777Z}).
\vskip 0.2cm
The star formation efficiency of a galaxy,   $\rm f_{\star}$,  can be expressed by \citep[][]{2019A&A...626A..56P},
\begin{equation} \label{eq:fstar}
    \rm{f_{\star} = \frac{M_{\star}}{M_{halo}}~\frac{\Omega_{m}}{\Omega_{b}}}
\end{equation}


For UGC~12591 If we take halo mass $\rm M_{halo} = \rm M_{200}$ as calculated in \S~\ref{subsec:btfr} and stellar mass $\rm M_{\star} = (1.6\pm0.1)\times10^{11} \Msun$ we get $\rm f_{\star} \sim 0.04\pm0.01$, for universal   cosmic baryon fraction $\frac{\Omega_{b}}{\Omega_{m}} = 0.174$.  This shows that in spite of a  very large halo mass
it is able to convert only a small fraction ($\rm 3 -5 \%$) of halo baryons to stars. On the contrary, the sample of most massive spirals with $\rm M_{\star} = 1-3\times10^{11} \Msun$ of \citet{2019A&A...626A..56P} yields $\rm f_{\star} \approx 0.3-1$ which shows their high  efficiency of star formation and their stellar baryon fraction is approaching the cosmological baryon fraction. 
Although a very interesting result,  it is still too early to tell if these unusual spirals of   \citet{2019A&A...626A..56P} have  experienced the same evolutionary and SFH as the majority of normal spirals.

UGC~12591  appears to be very  different  because of its very low $\rm f_{\star}$ and possibly most of the  cosmic baryons may reside  in the  dark matter halo, unable to  turn  into stars.  
Now if we insert  total baryonic mass (\S~\ref{subsec:btfr}) instead of stellar mass in Eq.~\ref{eq:fstar} assuming  all the baryons could have been converted to stars, we find $\rm f_{\star} \sim 0.15$, again highlighting  the stark  star formation deficiency of this galaxy. 
\citet[][]{2018ApJ...855L..24L} estimate the baryon budget of the hot circum galactic medium of six local isolated massive spiral galaxies, including UGC~12591, finding that a significant amount of the `missing baryons' are possibly stored beyond the virial radius of their dark matter halos and not
condensing into stars. Although such large repository of baryons beyond the virial radius is not yet found in any galaxy, if true,  this may
explain the insufficiency of the hot baryons inside virial radius to account for the expected cosmic baryons. 
These baryons may enable galaxies to continue forming stars steadily for long periods of time and account for missing baryons in galaxies in the local universe, but for most massive galaxies like UGC~12591  we find just the opposite. This finding is broadly consistent with the constraints from  abundance matching techniques to
determine the typical galaxy stellar mass at a given halo mass. The simulations show that  peak efficiency of $\rm f_{\star}$ is always fairly quite low,  at $\rm {\it z} = 0$ reaching a peak  of  about $\rm  20 - 30 \%$ for  typical $\rm L_{\star}$ galaxies  with halo mass $\rm  M_{200} \sim 10^{12} M_{\sun}$. Beyond the peak $\rm f_{\star}$ declines rapidly to  $\rm f_{\star} \sim 2 - 5 \%$ at $\rm  M_{200} \sim 10^{13} M_{\sun}$ \citep[][]{2010ApJ...717..379B,2019MNRAS.488.3143B,2013MNRAS.428.3121M}.  It is believed that AGN feedback is one of the main ingredients needed to bring this  decline in SFR, e.g. \citep[][]{2005Natur.433..604D,2017MNRAS.472..949B}.

To probe further the tension of  baryon budget of UGC~12591 and  standard cosmology we may compare it with the baryon fraction of  massive halos, including galaxy groups and clusters.
In  left panel Fig.~\ref{fig:fb_vrot} we plot the
fraction of all baryons, $\rm f_{b}$ (in stars + ISM + hot halo gas), and the right panel shows the baryon fraction only in hot halo phase, $\rm f_{b,hot}$, relative to the cosmic baryon fraction (the horizontal 
grey zone). Here the rotation velocity
$\rm v_{rot}$ is used as  proxy for the total mass of the halo. Now we see that the progressive increase of $\rm f_{b}$ at higher $\rm v_{rot}$, which approaches the cosmic value for cluster mass halos, indicating that most massive halos are nearly at closure for baryonic matter, compared to average galaxies such as the  Milky Way and even the most massive galaxies. In the baryon budget the stellar mass component begins to
dominate for galaxy mass halos while group and clusters are largely dominated by hot gas
phase. In spite of its large halo mass we  find a striking lack of baryon fraction in UGC~12591, even after including those contained in its X-ray halo (Fig.~\ref{fig:fb_vrot} right plot). The hot baryon fraction $\rm f_{b,hot}$ still remains significantly lower than inferred from X-ray observations of galaxy groups with similar $\rm v_{rot}$. 

On the contrary, the extremely massive, rapidly rotating, relativistic-jet-launching spiral galaxy J2345-0449 \citep[][]{2014ApJ...788..174B,2021A&A...654A...8N}
is an interesting and  rare exception, which has the baryon mass fraction within the virial radius 
($\sim 450$ kpc) of $\rm  f_{b} = 0.121\pm0.043$ \citep[][]{2021MNRAS.500.2503M,2015MNRAS.449.3527W}, similar  to the universal baryon fraction  of  group mass halos and  has  similar hot gas fraction (Fig.~\ref{fig:fb_vrot} right plot). NGC~1961 and NGC~6753 are also  close to it in terms of their baryonic budget. Although the halo mass of UGC~12591 is close to that of NGC~1961, NGC~6753 and J2345-0449, in terms of baryon content in hot halo, it falls short  of these other three remarkable galaxies.

Next we  compare the 0.5-2 keV band X-ray luminosity ($\rm L_{x}$) of all known   massive spiral galaxies with detected soft
X-ray halos.  We find that at $\rm L_{x} = 4.0\pm0.5 \times 10^{41}$ $\rm erg \, s^{-1}$ the luminosity of J2345-0449 is highest \citep[][]{2021MNRAS.500.2503M}, about 10 times that of  UGC~12591 \citep[][]{2012ApJ...755..107D}, NGC 6753 \citep[][]{2018ApJ...862....3B,2013ApJ...772...97B} and significantly
higher than that of NGC 1961 \citep[][]{2016MNRAS.455..227A}. The metal abundance of UGC~12591 halo is low at $\rm \sim 0.1 Z_{\sun}$,  similar to  those found for J2345-0449, NGC~1961 and NGC 6753. This  metallicity is consistent with baryon deficit,  low  metallicity coronal gas around massive galaxies shown in  EAGLE hydro simulations \citep[][]{2021MNRAS.502.2934K,2015MNRAS.446..521S}.  Although the
halo mass of UGC~12591, NGC~1961, NGC~6753 and J2345-0449 are similar, the higher X-ray luminosity and the larger  extent of J2345-0449 corona  (about 160 kpc, or 35 per cent of $\rm r_{200}$) is still intriguing. We point out  that J2345-0449  is the only one of those four galaxies which hosts   large scale   radio jets. Therefore,  possibly it is a consequence  of   powerful  Mpc scale radio jets and past supernovae  activity in this galaxy,  heating  \citep[][]{2021A&A...654A...8N}  and  expelling  the inner baryons upto a larger radius and  preventing  gas cooling in the halo, although more observational proof  is needed to demonstrate  this scenario robustly.

\section{Summary } 
\label{sec:summary}

In this paper we  analyse   the structure of an extremely fast rotating and massive, early-type or hybrid (S0/a) galaxy UGC~12591 by fitting PSF convolved 2D surface brightness models to HST/WFC3 images in the three bands of F606W (V), F814W (I) and F160W (H). In all of three bands a two component fit with \sersic~and exponential-disc function yields satisfactory results but adding one more component of PSF function optimizes the model further and improves the residuals. UGC~12591 is found to be a bulge dominated galaxy with \sersic~ index $\rm n$ ranging from $\rm 2.32-3.68$, effective bulge radius $\rm r_{e}$ of $\rm 7.35\arcsec-12.32\arcsec$ ($3.3 - 5.6$ kpc) and the disc scale length $\rm r_{s}$ of $\rm 41.56\arcsec-58.43\arcsec$ ($18.8 - 26.4$ kpc) in V, I and H bands, respectively. 

We have  calibrated the spectral energy distribution of the galaxy using two different packages, {\fontfamily{qcr}\selectfont CIGALE} and {\fontfamily{qcr}\selectfont MAGPHYS}. SED fitting by {\fontfamily{qcr}\selectfont CIGALE} gives total stellar mass of $(1.6\pm0.1)\times10^{11}\Msun$ with the ratio of old to young stellar mass as high as $(1.6\pm0.2)\times10^{5}$. The instantaneous star formation rate of $\rm 0.105\pm0.007 \Msun yr^{-1}$ and specific star formation rate, $ \rm (6.6\pm0.6) \times 10^{-13} \, yr^{-1}$
indicates  extreme quenching and near absence of recent star forming activity in the galaxy. We discussed where it stands compared to other massive galaxies in the same mass range and the possible effect of structural bulge and AGN activity in the  past in lowering the star formation. The SED fitting values of {\fontfamily{qcr}\selectfont CIGALE} are well consistent with the results of {\fontfamily{qcr}\selectfont MAGPHYS}, with a stellar mass of $9.12^{+1.38}_{-1.54}\times10^{10}\Msun$, SFR $\rm 0.210^{+0.037}_{-0.097} \Msun yr^{-1}$ and sSFR $ \rm 2.14^{+0.88}_{-0.94} \times 10^{-12} \, yr^{-1}$.

UGC~12591 possesses  both an extended ellipsoidal stellar bulge, and a giant equatorial disc of  gas and dust. The fraction of total infra-red dust luminosity contributed by the ISM dust is as high as $\rm 90\%$, with total dust luminosity $\rm L_{dust} = (0.50-2.75) \times 10^{10} \Lsun$, dominated by cold dust at  equilibrium temperature $\rm T_{C} = 16.64^{+1.67}_{-1.10} K$. 
 The total dust mass in ISM is estimated at $\rm 2.07^{+1.41}_{-1.05} \times 10^{8} \Msun$, much higher than normal spiral galaxies. 

Virial mass within $\rm R_{200}$ is calculated as $(2.47\pm0.89)\times10^{13}\Msun$ which is close to  previous  gravitational mass calculated from X-ray observations, 
assuming a hydrostatic equilibrium state for the hot X-ray corona. 
Having included the XMM-Newton detected X-ray halo mass in the baryonic budget, and accounting for all  ISM and stellar baryons, 
we find total baryonic mass of $\rm (6.46\pm0.20)\times10^{11} \Msun$ for UGC~12591. This  corresponds  to  total   baryon fraction $\rm \sim 0.03$, far less than cosmological baryonic closure value,  and  amounting to a baryon deficiency of $\rm \sim 85\%$ within the virial radius. However, within 80 kpc  (the  X-ray halo radius) the baryon fraction is $\rm \sim 0.15$, taking total gravitational mass as $4.4\times10^{12}\Msun$. This value is near the  cosmological baryon fraction which is $\rm \sim 0.17$. This implies that only a small fraction of these cosmic baryons reside in a galaxy scale  halo and about five  times more  baryonic  mass possibly resides in the extended  circum-galactic halo and  filaments  than what has been observed   so far in all wavelengths. 



On the Baryonic Tully Fisher relation  plot we show UGC~12591  falls  far below the well established linear relation of slope $\rm \sim4$.   All other known fast rotating and massive spirals like J2345-0449, NGC~1961, NGC~6753 and other
super massive spirals also 
follow  the same trend. After inclusion of hot X-ray halo baryons detected in J2345-0449, NGC~1961 and NGC~6753,  they all fall exactly  on  the  classic BTF relation   of slope $\rm \sim 3.7 - 4$. On the contrary, UGC~12591 still falls much  below of that line even after including  the hot halo gas mass extrapolated upto its virial radius. This could indicate a   deviation or breakdown  in the  self-similar scaling relation 
governing BTFR due to some non-gravitational processes such as AGN feedback, supernova blowout, bulge formation  or  a  longer radiative cooling time of baryons in CGM, due to some heating process. 

UGC~12591 is host to a supermassive black hole  of  mass $\rm (6.18\pm2.61)\times 10^{8} M_{\sun}$ which  at  the present moment is quiescent, i.e. we see neither large-scale ($\rm > 1 kpc$) radio jets nor is the  black hole contributing significantly to the  mid-IR SED,  ruling out the presence of strong  radiative feedback from a  bright AGN. A
compact central radio source  having 1.4 GHz luminosity of $\rm L_{1.4} = $ (3.01$\pm 0.92) \times 10^{32}$ W is found which can possibly be attributed to radio emission of an AGN. 
The possible role of intense AGN activity and energetic feedback of this black hole in the past, in quenching the star formation activity is discussed.

In this paper we showed that UGC~12591 is 
 one of the rare, most massive, fast rotating galaxies known in the local universe. It has  a large stellar mass  and  highly  star formation quenched state.  Because it is not obvious how this  galaxy  acquired such a range of extraordinary properties, it poses many challenges to
theoretical models. Further investigations via  multi-wavelength observations of it hold the key for understanding
the coevolution of black holes and massive disc galaxies, 
the strong quenching of star formation,  the complex interplay of dark-matter and baryonic physics taking place in  massive halos, and  the  role played by   AGN feedback and bulge formation in its evolution,  which  will help in explaining the unique characteristics  displayed  by this galaxy.

\section*{Acknowledgements}
The authors are grateful to the anonymous referee for their encouraging and constructive comments on the manuscript, that greatly helped us to improve its quality. 
SR gratefully acknowledges the support from IUCAA  under the student visiting program and funding by Indian Space Research Organisation (ISRO)  under `AstroSat Data Utilization'  project. 
JB acknowledges the support from Department of Physics and Electronics, CHRIST (Deemed to be University),
Bangalore. SD acknowledges the support from Department of Science and Technology (DST), New Delhi under the INSPIRE faculty  Scheme (sanctioned No: DST/INSPIRE/04/2015/000108).
MBP gratefully acknowledges the support from following funding schemes:  The Science and Engineering Research Board (SERB), New Delhi under the SERB ``SERB Research Scientists Scheme” Scheme and Indian Space Research Organisation (ISRO)  under `AstroSat Data Utilization'  project. MBP also acknowledge the support from IUCAA  Associateship  program.
This research has made use of the data from {\it HST} Archive.   Part of the reported results is based on observations made with the NASA/ESA Hubble Space Telescope, obtained from the Data Archive at the Space Telescope Science Institute, which is operated by the Association of Universities for Research in Astronomy,  Inc., under  NASA  contract  NAS 5-26555.  This research has made use of  NASA's  Astrophysics Data  System, and of the  NASA/IPAC  Extragalactic Database  (NED) which is operated by the Jet  Propulsion Laboratory, California Institute of Technology, under contract with the National Aeronautics and Space Administration.
We have used images and results from SDSS and funding for SDSS has been provided by the Alfred P. Sloan Foundation, the participating institutions, the National Science Foundation, and the U.S. Department of Energy’s Office of Science. This research has made use of the NASA/IPAC Extragalactic Database (NED) which is operated by the Jet Propulsion Laboratory, California Institute of Technology, under contract with the NASA.
Facilities: HST (ACS), SDSS.

\section*{Data Availability}
The data used in this paper are publicly available at https://hla.stsci.edu. 
The retrieved  data used in the analysis are tabulated in Table.~\ref{tab:hst_data_details} and Table.~\ref{tab:flux_data_ned}. The code used in reducing the HST observations is publicly available at https://users.obs.carnegiescience.edu/peng/work/galfit/galfit.html, http://www.iap.fr/magphys/, and https://cigale.lam.fr.



\bibliographystyle{mnras}
\bibliography{UGC12591} 

\begin{thebibliography}{}
\makeatletter
\relax
\def\mn@urlcharsother{\let\do\@makeother \do\$\do\&\do\#\do\^\do\_\do\%\do\~}
\def\mn@doi{\begingroup\mn@urlcharsother \@ifnextchar [ {\mn@doi@}
  {\mn@doi@[]}}
\def\mn@doi@[#1]#2{\def\@tempa{#1}\ifx\@tempa\@empty \href
  {http://dx.doi.org/#2} {doi:#2}\else \href {http://dx.doi.org/#2} {#1}\fi
  \endgroup}
\def\mn@eprint#1#2{\mn@eprint@#1:#2::\@nil}
\def\mn@eprint@arXiv#1{\href {http://arxiv.org/abs/#1} {{\tt arXiv:#1}}}
\def\mn@eprint@dblp#1{\href {http://dblp.uni-trier.de/rec/bibtex/#1.xml}
  {dblp:#1}}
\def\mn@eprint@#1:#2:#3:#4\@nil{\def\@tempa {#1}\def\@tempb {#2}\def\@tempc
  {#3}\ifx \@tempc \@empty \let \@tempc \@tempb \let \@tempb \@tempa \fi \ifx
  \@tempb \@empty \def\@tempb {arXiv}\fi \@ifundefined
  {mn@eprint@\@tempb}{\@tempb:\@tempc}{\expandafter \expandafter \csname
  mn@eprint@\@tempb\endcsname \expandafter{\@tempc}}}

\bibitem[\protect\citeauthoryear{{Anderson} \& {Bregman}}{{Anderson} \&
  {Bregman}}{2011}]{2011ApJ...737...22A}
{Anderson} M.~E.,  {Bregman} J.~N.,  2011, \mn@doi [\apj]
  {10.1088/0004-637X/737/1/22}, \href
  {https://ui.adsabs.harvard.edu/abs/2011ApJ...737...22A} {737, 22}

\bibitem[\protect\citeauthoryear{{Anderson}, {Churazov}  \&
  {Bregman}}{{Anderson} et~al.}{2016}]{2016MNRAS.455..227A}
{Anderson} M.~E.,  {Churazov} E.,   {Bregman} J.~N.,  2016, \mn@doi [\mnras]
  {10.1093/mnras/stv2314}, \href
  {https://ui.adsabs.harvard.edu/abs/2016MNRAS.455..227A} {455, 227}

\bibitem[\protect\citeauthoryear{{Bagchi} et~al.,}{{Bagchi}
  et~al.}{2014}]{2014ApJ...788..174B}
{Bagchi} J.,  et~al., 2014, \mn@doi [\apj] {10.1088/0004-637X/788/2/174}, \href
  {https://ui.adsabs.harvard.edu/abs/2014ApJ...788..174B} {788, 174}

\bibitem[\protect\citeauthoryear{{Bandara}, {Crampton}  \& {Simard}}{{Bandara}
  et~al.}{2009}]{2009ApJ...704.1135B}
{Bandara} K.,  {Crampton} D.,   {Simard} L.,  2009, \mn@doi [\apj]
  {10.1088/0004-637X/704/2/1135}, \href
  {https://ui.adsabs.harvard.edu/abs/2009ApJ...704.1135B} {704, 1135}

\bibitem[\protect\citeauthoryear{{Beckmann} et~al.,}{{Beckmann}
  et~al.}{2017}]{2017MNRAS.472..949B}
{Beckmann} R.~S.,  et~al., 2017, \mn@doi [\mnras] {10.1093/mnras/stx1831},
  \href {https://ui.adsabs.harvard.edu/abs/2017MNRAS.472..949B} {472, 949}

\bibitem[\protect\citeauthoryear{{Behroozi}, {Conroy}  \&
  {Wechsler}}{{Behroozi} et~al.}{2010}]{2010ApJ...717..379B}
{Behroozi} P.~S.,  {Conroy} C.,   {Wechsler} R.~H.,  2010, \mn@doi [\apj]
  {10.1088/0004-637X/717/1/379}, \href
  {https://ui.adsabs.harvard.edu/abs/2010ApJ...717..379B} {717, 379}

\bibitem[\protect\citeauthoryear{{Behroozi}, {Wechsler}, {Hearin}  \&
  {Conroy}}{{Behroozi} et~al.}{2019}]{2019MNRAS.488.3143B}
{Behroozi} P.,  {Wechsler} R.~H.,  {Hearin} A.~P.,   {Conroy} C.,  2019,
  \mn@doi [\mnras] {10.1093/mnras/stz1182}, \href
  {https://ui.adsabs.harvard.edu/abs/2019MNRAS.488.3143B} {488, 3143}

\bibitem[\protect\citeauthoryear{{Bell} \& {de Jong}}{{Bell} \& {de
  Jong}}{2001}]{2001ApJ...550..212B}
{Bell} E.~F.,  {de Jong} R.~S.,  2001, \mn@doi [\apj] {10.1086/319728}, \href
  {https://ui.adsabs.harvard.edu/abs/2001ApJ...550..212B} {550, 212}

\bibitem[\protect\citeauthoryear{{Bendo} et~al.,}{{Bendo}
  et~al.}{2015}]{2015MNRAS.448..135B}
{Bendo} G.~J.,  et~al., 2015, \mn@doi [\mnras] {10.1093/mnras/stu1841}, \href
  {https://ui.adsabs.harvard.edu/abs/2015MNRAS.448..135B} {448, 135}

\bibitem[\protect\citeauthoryear{{Bertin} \& {Arnouts}}{{Bertin} \&
  {Arnouts}}{1996}]{1996A&AS..117..393B}
{Bertin} E.,  {Arnouts} S.,  1996, \mn@doi [\aaps] {10.1051/aas:1996164}, \href
  {https://ui.adsabs.harvard.edu/abs/1996A&AS..117..393B} {117, 393}

\bibitem[\protect\citeauthoryear{{Bogd{\'a}n} et~al.,}{{Bogd{\'a}n}
  et~al.}{2013}]{2013ApJ...772...97B}
{Bogd{\'a}n} {\'A}.,  et~al., 2013, \mn@doi [\apj]
  {10.1088/0004-637X/772/2/97}, \href
  {https://ui.adsabs.harvard.edu/abs/2013ApJ...772...97B} {772, 97}

\bibitem[\protect\citeauthoryear{{Boquien}, {Burgarella}, {Roehlly}, {Buat},
  {Ciesla}, {Corre}, {Inoue}  \& {Salas}}{{Boquien}
  et~al.}{2019}]{2019A&A...622A.103B}
{Boquien} M.,  {Burgarella} D.,  {Roehlly} Y.,  {Buat} V.,  {Ciesla} L.,
  {Corre} D.,  {Inoue} A.~K.,   {Salas} H.,  2019, \mn@doi [\aap]
  {10.1051/0004-6361/201834156}, \href
  {https://ui.adsabs.harvard.edu/abs/2019A&A...622A.103B} {622, A103}

\bibitem[\protect\citeauthoryear{{Bregman}}{{Bregman}}{2007}]{2007ARA&A..45..221B}
{Bregman} J.~N.,  2007, \mn@doi [\araa]
  {10.1146/annurev.astro.45.051806.110619}, \href
  {https://ui.adsabs.harvard.edu/abs/2007ARA&A..45..221B} {45, 221}

\bibitem[\protect\citeauthoryear{{Bregman}, {Anderson}, {Miller},
  {Hodges-Kluck}, {Dai}, {Li}, {Li}  \& {Qu}}{{Bregman}
  et~al.}{2018}]{2018ApJ...862....3B}
{Bregman} J.~N.,  {Anderson} M.~E.,  {Miller} M.~J.,  {Hodges-Kluck} E.,  {Dai}
  X.,  {Li} J.-T.,  {Li} Y.,   {Qu} Z.,  2018, \mn@doi [\apj]
  {10.3847/1538-4357/aacafe}, \href
  {https://ui.adsabs.harvard.edu/abs/2018ApJ...862....3B} {862, 3}

\bibitem[\protect\citeauthoryear{{Bruzual} \& {Charlot}}{{Bruzual} \&
  {Charlot}}{2003}]{2003MNRAS.344.1000B}
{Bruzual} G.,  {Charlot} S.,  2003, \mn@doi [\mnras]
  {10.1046/j.1365-8711.2003.06897.x}, \href
  {https://ui.adsabs.harvard.edu/abs/2003MNRAS.344.1000B} {344, 1000}

\bibitem[\protect\citeauthoryear{{Burgarella}, {Buat}  \&
  {Iglesias-P{\'a}ramo}}{{Burgarella} et~al.}{2005}]{2005MNRAS.360.1413B}
{Burgarella} D.,  {Buat} V.,   {Iglesias-P{\'a}ramo} J.,  2005, \mn@doi
  [\mnras] {10.1111/j.1365-2966.2005.09131.x}, \href
  {https://ui.adsabs.harvard.edu/abs/2005MNRAS.360.1413B} {360, 1413}

\bibitem[\protect\citeauthoryear{{Condon}, {Cotton}, {Greisen}, {Yin},
  {Perley}, {Taylor}  \& {Broderick}}{{Condon}
  et~al.}{1998}]{1998AJ....115.1693C}
{Condon} J.~J.,  {Cotton} W.~D.,  {Greisen} E.~W.,  {Yin} Q.~F.,  {Perley}
  R.~A.,  {Taylor} G.~B.,   {Broderick} J.~J.,  1998, \mn@doi [\aj]
  {10.1086/300337}, \href
  {https://ui.adsabs.harvard.edu/abs/1998AJ....115.1693C} {115, 1693}

\bibitem[\protect\citeauthoryear{{Condon}, {Cotton}  \& {Broderick}}{{Condon}
  et~al.}{2002}]{2002AJ....124..675C}
{Condon} J.~J.,  {Cotton} W.~D.,   {Broderick} J.~J.,  2002, \mn@doi [\aj]
  {10.1086/341650}, \href
  {https://ui.adsabs.harvard.edu/abs/2002AJ....124..675C} {124, 675}

\bibitem[\protect\citeauthoryear{{Courtois}, {Zaritsky}, {Sorce}  \&
  {Pomar{\`e}de}}{{Courtois} et~al.}{2015}]{2015MNRAS.448.1767C}
{Courtois} H.~M.,  {Zaritsky} D.,  {Sorce} J.~G.,   {Pomar{\`e}de} D.,  2015,
  \mn@doi [\mnras] {10.1093/mnras/stv071}, \href
  {https://ui.adsabs.harvard.edu/abs/2015MNRAS.448.1767C} {448, 1767}

\bibitem[\protect\citeauthoryear{{Cutri} et~al.,}{{Cutri}
  et~al.}{2003}]{2003yCat.2246....0C}
{Cutri} R.~M.,  et~al., 2003, VizieR Online Data Catalog, \href
  {https://ui.adsabs.harvard.edu/abs/2003yCat.2246....0C} {p. II/246}

\bibitem[\protect\citeauthoryear{{Cutri} et~al.,}{{Cutri}
  et~al.}{2013}]{2013wise.rept....1C}
{Cutri} R.~M.,  et~al., 2013, {Explanatory Supplement to the AllWISE Data
  Release Products}, Explanatory Supplement to the AllWISE Data Release
  Products, by R. M. Cutri et al.

\bibitem[\protect\citeauthoryear{{Dai}, {Bregman}, {Kochanek}  \&
  {Rasia}}{{Dai} et~al.}{2010}]{2010ApJ...719..119D}
{Dai} X.,  {Bregman} J.~N.,  {Kochanek} C.~S.,   {Rasia} E.,  2010, \mn@doi
  [\apj] {10.1088/0004-637X/719/1/119}, \href
  {https://ui.adsabs.harvard.edu/abs/2010ApJ...719..119D} {719, 119}

\bibitem[\protect\citeauthoryear{{Dai}, {Anderson}, {Bregman}  \&
  {Miller}}{{Dai} et~al.}{2012}]{2012ApJ...755..107D}
{Dai} X.,  {Anderson} M.~E.,  {Bregman} J.~N.,   {Miller} J.~M.,  2012, \mn@doi
  [\apj] {10.1088/0004-637X/755/2/107}, \href
  {https://ui.adsabs.harvard.edu/abs/2012ApJ...755..107D} {755, 107}

\bibitem[\protect\citeauthoryear{{Dale}, {Helou}, {Magdis}, {Armus},
  {D{\'\i}az-Santos}  \& {Shi}}{{Dale} et~al.}{2014}]{2014ApJ...784...83D}
{Dale} D.~A.,  {Helou} G.,  {Magdis} G.~E.,  {Armus} L.,  {D{\'\i}az-Santos}
  T.,   {Shi} Y.,  2014, \mn@doi [\apj] {10.1088/0004-637X/784/1/83}, \href
  {https://ui.adsabs.harvard.edu/abs/2014ApJ...784...83D} {784, 83}

\bibitem[\protect\citeauthoryear{{Dekel} \& {Burkert}}{{Dekel} \&
  {Burkert}}{2014}]{2014MNRAS.438.1870D}
{Dekel} A.,  {Burkert} A.,  2014, \mn@doi [\mnras] {10.1093/mnras/stt2331},
  \href {https://ui.adsabs.harvard.edu/abs/2014MNRAS.438.1870D} {438, 1870}

\bibitem[\protect\citeauthoryear{{Di Matteo}, {Springel}  \& {Hernquist}}{{Di
  Matteo} et~al.}{2005}]{2005Natur.433..604D}
{Di Matteo} T.,  {Springel} V.,   {Hernquist} L.,  2005, \mn@doi [\nat]
  {10.1038/nature03335}, \href
  {https://ui.adsabs.harvard.edu/abs/2005Natur.433..604D} {433, 604}

\bibitem[\protect\citeauthoryear{{Doyle} \& {Drinkwater}}{{Doyle} \&
  {Drinkwater}}{2006}]{2006MNRAS.372..977D}
{Doyle} M.~T.,  {Drinkwater} M.~J.,  2006, \mn@doi [\mnras]
  {10.1111/j.1365-2966.2006.10931.x}, \href
  {https://ui.adsabs.harvard.edu/abs/2006MNRAS.372..977D} {372, 977}

\bibitem[\protect\citeauthoryear{{Elbaz} et~al.,}{{Elbaz}
  et~al.}{2007}]{2007A&A...468...33E}
{Elbaz} D.,  et~al., 2007, \mn@doi [\aap] {10.1051/0004-6361:20077525}, \href
  {https://ui.adsabs.harvard.edu/abs/2007A&A...468...33E} {468, 33}

\bibitem[\protect\citeauthoryear{{Fabian}}{{Fabian}}{2012}]{2012ARA&A..50..455F}
{Fabian} A.~C.,  2012, \mn@doi [\araa] {10.1146/annurev-astro-081811-125521},
  \href {https://ui.adsabs.harvard.edu/abs/2012ARA&A..50..455F} {50, 455}

\bibitem[\protect\citeauthoryear{{Federrath} \& {Klessen}}{{Federrath} \&
  {Klessen}}{2012}]{2012ApJ...761..156F}
{Federrath} C.,  {Klessen} R.~S.,  2012, \mn@doi [\apj]
  {10.1088/0004-637X/761/2/156}, \href
  {https://ui.adsabs.harvard.edu/abs/2012ApJ...761..156F} {761, 156}

\bibitem[\protect\citeauthoryear{{Fisher} \& {Drory}}{{Fisher} \&
  {Drory}}{2008}]{2008AJ....136..773F}
{Fisher} D.~B.,  {Drory} N.,  2008, \mn@doi [\aj]
  {10.1088/0004-6256/136/2/773}, \href
  {https://ui.adsabs.harvard.edu/abs/2008AJ....136..773F} {136, 773}

\bibitem[\protect\citeauthoryear{{Fukugita} \& {Peebles}}{{Fukugita} \&
  {Peebles}}{2006}]{2006ApJ...639..590F}
{Fukugita} M.,  {Peebles} P.~J.~E.,  2006, \mn@doi [\apj] {10.1086/499556},
  \href {https://ui.adsabs.harvard.edu/abs/2006ApJ...639..590F} {639, 590}

\bibitem[\protect\citeauthoryear{{Gebhardt} et~al.,}{{Gebhardt}
  et~al.}{2000}]{2000ApJ...539L..13G}
{Gebhardt} K.,  et~al., 2000, \mn@doi [\apjl] {10.1086/312840}, \href
  {https://ui.adsabs.harvard.edu/abs/2000ApJ...539L..13G} {539, L13}

\bibitem[\protect\citeauthoryear{{Geha}, {Blanton}, {Masjedi}  \&
  {West}}{{Geha} et~al.}{2006}]{2006ApJ...653..240G}
{Geha} M.,  {Blanton} M.~R.,  {Masjedi} M.,   {West} A.~A.,  2006, \mn@doi
  [\apj] {10.1086/508604}, \href
  {https://ui.adsabs.harvard.edu/abs/2006ApJ...653..240G} {653, 240}

\bibitem[\protect\citeauthoryear{{Giovanelli}, {Haynes}, {Rubin}  \&
  {Ford}}{{Giovanelli} et~al.}{1986}]{1986ApJ...301L...7G}
{Giovanelli} R.,  {Haynes} M.~P.,  {Rubin} V.~C.,   {Ford} W.~K. J.,  1986,
  \mn@doi [\apjl] {10.1086/184613}, \href
  {https://ui.adsabs.harvard.edu/abs/1986ApJ...301L...7G} {301, L7}

\bibitem[\protect\citeauthoryear{{G{\"u}ltekin} et~al.,}{{G{\"u}ltekin}
  et~al.}{2009}]{2009ApJ...698..198G}
{G{\"u}ltekin} K.,  et~al., 2009, \mn@doi [\apj] {10.1088/0004-637X/698/1/198},
  \href {https://ui.adsabs.harvard.edu/abs/2009ApJ...698..198G} {698, 198}

\bibitem[\protect\citeauthoryear{{H{\"a}ring} \& {Rix}}{{H{\"a}ring} \&
  {Rix}}{2004}]{2004ApJ...604L..89H}
{H{\"a}ring} N.,  {Rix} H.-W.,  2004, \mn@doi [\apjl] {10.1086/383567}, \href
  {https://ui.adsabs.harvard.edu/abs/2004ApJ...604L..89H} {604, L89}

\bibitem[\protect\citeauthoryear{{Henden}, {Levine}, {Terrell}  \&
  {Welch}}{{Henden} et~al.}{2015}]{2015AAS...22533616H}
{Henden} A.~A.,  {Levine} S.,  {Terrell} D.,   {Welch} D.~L.,  2015, in
  American Astronomical Society Meeting Abstracts \#225. p. 336.16

\bibitem[\protect\citeauthoryear{{Hoopes} et~al.,}{{Hoopes}
  et~al.}{2007}]{2007ApJS..173..441H}
{Hoopes} C.~G.,  et~al., 2007, \mn@doi [\apjs] {10.1086/516644}, \href
  {https://ui.adsabs.harvard.edu/abs/2007ApJS..173..441H} {173, 441}

\bibitem[\protect\citeauthoryear{{Inoue}}{{Inoue}}{2011}]{2011MNRAS.415.2920I}
{Inoue} A.~K.,  2011, \mn@doi [\mnras] {10.1111/j.1365-2966.2011.18906.x},
  \href {https://ui.adsabs.harvard.edu/abs/2011MNRAS.415.2920I} {415, 2920}

\bibitem[\protect\citeauthoryear{{Intema}, {Jagannathan}, {Mooley}  \&
  {Frail}}{{Intema} et~al.}{2017}]{2017A&A...598A..78I}
{Intema} H.~T.,  {Jagannathan} P.,  {Mooley} K.~P.,   {Frail} D.~A.,  2017,
  \mn@doi [\aap] {10.1051/0004-6361/201628536}, \href
  {https://ui.adsabs.harvard.edu/abs/2017A&A...598A..78I} {598, A78}

\bibitem[\protect\citeauthoryear{{Jarrett} et~al.,}{{Jarrett}
  et~al.}{2013}]{2013AJ....145....6J}
{Jarrett} T.~H.,  et~al., 2013, \mn@doi [\aj] {10.1088/0004-6256/145/1/6},
  \href {https://ui.adsabs.harvard.edu/abs/2013AJ....145....6J} {145, 6}

\bibitem[\protect\citeauthoryear{{Jurlin} et~al.,}{{Jurlin}
  et~al.}{2020}]{2020A&A...638A..34J}
{Jurlin} N.,  et~al., 2020, \mn@doi [\aap] {10.1051/0004-6361/201936955}, \href
  {https://ui.adsabs.harvard.edu/abs/2020A&A...638A..34J} {638, A34}

\bibitem[\protect\citeauthoryear{{Kalinova}, {Colombo}, {S{\'a}nchez},
  {Kodaira}, {Garc{\'\i}a-Benito}, {Gonz{\'a}lez Delgado}, {Rosolowsky}  \&
  {Lacerda}}{{Kalinova} et~al.}{2021}]{2021A&A...648A..64K}
{Kalinova} V.,  {Colombo} D.,  {S{\'a}nchez} S.~F.,  {Kodaira} K.,
  {Garc{\'\i}a-Benito} R.,  {Gonz{\'a}lez Delgado} R.,  {Rosolowsky} E.,
  {Lacerda} E.~A.~D.,  2021, \mn@doi [\aap] {10.1051/0004-6361/202039896},
  \href {https://ui.adsabs.harvard.edu/abs/2021A&A...648A..64K} {648, A64}

\bibitem[\protect\citeauthoryear{{Kannan}, {Macci{\`o}}, {Fontanot}, {Moster},
  {Karman}  \& {Somerville}}{{Kannan} et~al.}{2015}]{2015MNRAS.452.4347K}
{Kannan} R.,  {Macci{\`o}} A.~V.,  {Fontanot} F.,  {Moster} B.~P.,  {Karman}
  W.,   {Somerville} R.~S.,  2015, \mn@doi [\mnras] {10.1093/mnras/stv1633},
  \href {https://ui.adsabs.harvard.edu/abs/2015MNRAS.452.4347K} {452, 4347}

\bibitem[\protect\citeauthoryear{{Kelly}, {Jenkins}  \& {Frenk}}{{Kelly}
  et~al.}{2021}]{2021MNRAS.502.2934K}
{Kelly} A.~J.,  {Jenkins} A.,   {Frenk} C.~S.,  2021, \mn@doi [\mnras]
  {10.1093/mnras/stab255}, \href
  {https://ui.adsabs.harvard.edu/abs/2021MNRAS.502.2934K} {502, 2934}

\bibitem[\protect\citeauthoryear{{Kennicutt}}{{Kennicutt}}{1998}]{Kennicutt98}
{Kennicutt} Robert~C. J.,  1998, \mn@doi [\apj] {10.1086/305588}, \href
  {https://ui.adsabs.harvard.edu/abs/1998ApJ...498..541K} {498, 541}

\bibitem[\protect\citeauthoryear{{Khoperskov}, {Haywood}, {Di Matteo},
  {Lehnert}  \& {Combes}}{{Khoperskov} et~al.}{2018}]{2018A&A...609A..60K}
{Khoperskov} S.,  {Haywood} M.,  {Di Matteo} P.,  {Lehnert} M.~D.,   {Combes}
  F.,  2018, \mn@doi [\aap] {10.1051/0004-6361/201731211}, \href
  {https://ui.adsabs.harvard.edu/abs/2018A&A...609A..60K} {609, A60}

\bibitem[\protect\citeauthoryear{{Knapp}, {Guhathakurta}, {Kim}  \&
  {Jura}}{{Knapp} et~al.}{1989}]{1989ApJS...70..329K}
{Knapp} G.~R.,  {Guhathakurta} P.,  {Kim} D.-W.,   {Jura} M.~A.,  1989, \mn@doi
  [\apjs] {10.1086/191342}, \href
  {https://ui.adsabs.harvard.edu/abs/1989ApJS...70..329K} {70, 329}

\bibitem[\protect\citeauthoryear{{Komatsu} et~al.,}{{Komatsu}
  et~al.}{2009}]{2009ApJS..180..330K}
{Komatsu} E.,  et~al., 2009, \mn@doi [\apjs] {10.1088/0067-0049/180/2/330},
  \href {https://ui.adsabs.harvard.edu/abs/2009ApJS..180..330K} {180, 330}

\bibitem[\protect\citeauthoryear{{Kormendy} \& {Ho}}{{Kormendy} \&
  {Ho}}{2013}]{2013ARA&A..51..511K}
{Kormendy} J.,  {Ho} L.~C.,  2013, \mn@doi [\araa]
  {10.1146/annurev-astro-082708-101811}, \href
  {https://ui.adsabs.harvard.edu/abs/2013ARA&A..51..511K} {51, 511}

\bibitem[\protect\citeauthoryear{{Lanz}, {Ogle}, {Alatalo}  \&
  {Appleton}}{{Lanz} et~al.}{2016}]{2016ApJ...826...29L}
{Lanz} L.,  {Ogle} P.~M.,  {Alatalo} K.,   {Appleton} P.~N.,  2016, \mn@doi
  [\apj] {10.3847/0004-637X/826/1/29}, \href
  {https://ui.adsabs.harvard.edu/abs/2016ApJ...826...29L} {826, 29}

\bibitem[\protect\citeauthoryear{{Lelli}, {McGaugh}  \& {Schombert}}{{Lelli}
  et~al.}{2016a}]{2016AJ....152..157L}
{Lelli} F.,  {McGaugh} S.~S.,   {Schombert} J.~M.,  2016a, \mn@doi [\aj]
  {10.3847/0004-6256/152/6/157}, \href
  {https://ui.adsabs.harvard.edu/abs/2016AJ....152..157L} {152, 157}

\bibitem[\protect\citeauthoryear{{Lelli}, {McGaugh}  \& {Schombert}}{{Lelli}
  et~al.}{2016b}]{2016ApJ...816L..14L}
{Lelli} F.,  {McGaugh} S.~S.,   {Schombert} J.~M.,  2016b, \mn@doi [\apjl]
  {10.3847/2041-8205/816/1/L14}, \href
  {https://ui.adsabs.harvard.edu/abs/2016ApJ...816L..14L} {816, L14}

\bibitem[\protect\citeauthoryear{{Li}, {Crain}  \& {Wang}}{{Li}
  et~al.}{2014}]{2014MNRAS.440..859L}
{Li} J.-T.,  {Crain} R.~A.,   {Wang} Q.~D.,  2014, \mn@doi [\mnras]
  {10.1093/mnras/stu329}, \href
  {https://ui.adsabs.harvard.edu/abs/2014MNRAS.440..859L} {440, 859}

\bibitem[\protect\citeauthoryear{{Li}, {Bregman}, {Wang}, {Crain}, {Anderson}
  \& {Zhang}}{{Li} et~al.}{2017}]{2017ApJS..233...20L}
{Li} J.-T.,  {Bregman} J.~N.,  {Wang} Q.~D.,  {Crain} R.~A.,  {Anderson} M.~E.,
    {Zhang} S.,  2017, \mn@doi [\apjs] {10.3847/1538-4365/aa96fc}, \href
  {https://ui.adsabs.harvard.edu/abs/2017ApJS..233...20L} {233, 20}

\bibitem[\protect\citeauthoryear{{Li}, {Bregman}, {Wang}, {Crain}  \&
  {Anderson}}{{Li} et~al.}{2018}]{2018ApJ...855L..24L}
{Li} J.-T.,  {Bregman} J.~N.,  {Wang} Q.~D.,  {Crain} R.~A.,   {Anderson}
  M.~E.,  2018, \mn@doi [\apjl] {10.3847/2041-8213/aab2af}, \href
  {https://ui.adsabs.harvard.edu/abs/2018ApJ...855L..24L} {855, L24}

\bibitem[\protect\citeauthoryear{{Lynden-Bell}}{{Lynden-Bell}}{1969}]{1969Natur.223..690L}
{Lynden-Bell} D.,  1969, \mn@doi [\nat] {10.1038/223690a0}, \href
  {https://ui.adsabs.harvard.edu/abs/1969Natur.223..690L} {223, 690}

\bibitem[\protect\citeauthoryear{{MacArthur}, {Courteau}  \&
  {Holtzman}}{{MacArthur} et~al.}{2003}]{2003ApJ...582..689M}
{MacArthur} L.~A.,  {Courteau} S.,   {Holtzman} J.~A.,  2003, \mn@doi [\apj]
  {10.1086/344506}, \href
  {https://ui.adsabs.harvard.edu/abs/2003ApJ...582..689M} {582, 689}

\bibitem[\protect\citeauthoryear{{Madau} \& {Dickinson}}{{Madau} \&
  {Dickinson}}{2014}]{2014ARA&A..52..415M}
{Madau} P.,  {Dickinson} M.,  2014, \mn@doi [\araa]
  {10.1146/annurev-astro-081811-125615}, \href
  {https://ui.adsabs.harvard.edu/abs/2014ARA&A..52..415M} {52, 415}

\bibitem[\protect\citeauthoryear{{Maddox}, {Hess}, {Obreschkow}, {Jarvis}  \&
  {Blyth}}{{Maddox} et~al.}{2015}]{2015MNRAS.447.1610M}
{Maddox} N.,  {Hess} K.~M.,  {Obreschkow} D.,  {Jarvis} M.~J.,   {Blyth} S.~L.,
   2015, \mn@doi [\mnras] {10.1093/mnras/stu2532}, \href
  {https://ui.adsabs.harvard.edu/abs/2015MNRAS.447.1610M} {447, 1610}

\bibitem[\protect\citeauthoryear{{Magorrian} et~al.,}{{Magorrian}
  et~al.}{1998}]{1998AJ....115.2285M}
{Magorrian} J.,  et~al., 1998, \mn@doi [\aj] {10.1086/300353}, \href
  {https://ui.adsabs.harvard.edu/abs/1998AJ....115.2285M} {115, 2285}

\bibitem[\protect\citeauthoryear{{Makarov}, {Prugniel}, {Terekhova}, {Courtois}
   \& {Vauglin}}{{Makarov} et~al.}{2014}]{2014A&A...570A..13M}
{Makarov} D.,  {Prugniel} P.,  {Terekhova} N.,  {Courtois} H.,   {Vauglin} I.,
  2014, \mn@doi [\aap] {10.1051/0004-6361/201423496}, \href
  {http://adsabs.harvard.edu/abs/2014A%26A...570A..13M} {570, A13}

\bibitem[\protect\citeauthoryear{{Mandal}, {Mukherjee}, {Federrath},
  {Nesvadba}, {Bicknell}, {Wagner}  \& {Meenakshi}}{{Mandal}
  et~al.}{2021}]{2021MNRAS.508.4738M}
{Mandal} A.,  {Mukherjee} D.,  {Federrath} C.,  {Nesvadba} N. P.~H.,
  {Bicknell} G.~V.,  {Wagner} A.~Y.,   {Meenakshi} M.,  2021, \mn@doi [\mnras]
  {10.1093/mnras/stab2822}, \href
  {https://ui.adsabs.harvard.edu/abs/2021MNRAS.508.4738M} {508, 4738}

\bibitem[\protect\citeauthoryear{{Marconi} \& {Hunt}}{{Marconi} \&
  {Hunt}}{2003}]{2003ApJ...589L..21M}
{Marconi} A.,  {Hunt} L.~K.,  2003, \mn@doi [\apjl] {10.1086/375804}, \href
  {https://ui.adsabs.harvard.edu/abs/2003ApJ...589L..21M} {589, L21}

\bibitem[\protect\citeauthoryear{{Martig}, {Bournaud}, {Teyssier}  \&
  {Dekel}}{{Martig} et~al.}{2009}]{2009ApJ...707..250M}
{Martig} M.,  {Bournaud} F.,  {Teyssier} R.,   {Dekel} A.,  2009, \mn@doi
  [\apj] {10.1088/0004-637X/707/1/250}, \href
  {https://ui.adsabs.harvard.edu/abs/2009ApJ...707..250M} {707, 250}

\bibitem[\protect\citeauthoryear{{Martig} et~al.,}{{Martig}
  et~al.}{2013}]{2013MNRAS.432.1914M}
{Martig} M.,  et~al., 2013, \mn@doi [\mnras] {10.1093/mnras/sts594}, \href
  {https://ui.adsabs.harvard.edu/abs/2013MNRAS.432.1914M} {432, 1914}

\bibitem[\protect\citeauthoryear{{Mart{\'\i}n-Navarro}, {Brodie}, {Romanowsky},
  {Ruiz-Lara}  \& {van de Ven}}{{Mart{\'\i}n-Navarro}
  et~al.}{2018}]{2018Natur.553..307M}
{Mart{\'\i}n-Navarro} I.,  {Brodie} J.~P.,  {Romanowsky} A.~J.,  {Ruiz-Lara}
  T.,   {van de Ven} G.,  2018, \mn@doi [\nat] {10.1038/nature24999}, \href
  {https://ui.adsabs.harvard.edu/abs/2018Natur.553..307M} {553, 307}

\bibitem[\protect\citeauthoryear{{McGaugh}}{{McGaugh}}{2005}]{2005ApJ...632..859M}
{McGaugh} S.~S.,  2005, \mn@doi [\apj] {10.1086/432968}, \href
  {https://ui.adsabs.harvard.edu/abs/2005ApJ...632..859M} {632, 859}

\bibitem[\protect\citeauthoryear{McGaugh}{McGaugh}{2012}]{McGaugh_2012}
McGaugh S.~S.,  2012, \mn@doi [The Astronomical Journal]
  {10.1088/0004-6256/143/2/40}, 143, 40

\bibitem[\protect\citeauthoryear{{McGaugh}, {Schombert}, {Bothun}  \& {de
  Blok}}{{McGaugh} et~al.}{2000}]{2000ApJ...533L..99M}
{McGaugh} S.~S.,  {Schombert} J.~M.,  {Bothun} G.~D.,   {de Blok} W.~J.~G.,
  2000, \mn@doi [\apjl] {10.1086/312628}, \href
  {https://ui.adsabs.harvard.edu/abs/2000ApJ...533L..99M} {533, L99}

\bibitem[\protect\citeauthoryear{{Meiksin}}{{Meiksin}}{2006}]{2006MNRAS.365..807M}
{Meiksin} A.,  2006, \mn@doi [\mnras] {10.1111/j.1365-2966.2005.09756.x}, \href
  {https://ui.adsabs.harvard.edu/abs/2006MNRAS.365..807M} {365, 807}

\bibitem[\protect\citeauthoryear{{Miller} \& {Bregman}}{{Miller} \&
  {Bregman}}{2015}]{2015ApJ...800...14M}
{Miller} M.~J.,  {Bregman} J.~N.,  2015, \mn@doi [\apj]
  {10.1088/0004-637X/800/1/14}, \href
  {https://ui.adsabs.harvard.edu/abs/2015ApJ...800...14M} {800, 14}

\bibitem[\protect\citeauthoryear{{Mirakhor} et~al.,}{{Mirakhor}
  et~al.}{2021}]{2021MNRAS.500.2503M}
{Mirakhor} M.~S.,  et~al., 2021, \mn@doi [\mnras] {10.1093/mnras/staa3404},
  \href {https://ui.adsabs.harvard.edu/abs/2021MNRAS.500.2503M} {500, 2503}

\bibitem[\protect\citeauthoryear{{M{\"o}llenhoff}}{{M{\"o}llenhoff}}{2004}]{2004A&A...415...63M}
{M{\"o}llenhoff} C.,  2004, \mn@doi [\aap] {10.1051/0004-6361:20034122}, \href
  {https://ui.adsabs.harvard.edu/abs/2004A&A...415...63M} {415, 63}

\bibitem[\protect\citeauthoryear{{Morganti}, {Tadhunter}  \&
  {Oosterloo}}{{Morganti} et~al.}{2005}]{2005A&A...444L...9M}
{Morganti} R.,  {Tadhunter} C.~N.,   {Oosterloo} T.~A.,  2005, \mn@doi [\aap]
  {10.1051/0004-6361:200500197}, \href
  {https://ui.adsabs.harvard.edu/abs/2005A&A...444L...9M} {444, L9}

\bibitem[\protect\citeauthoryear{{Moster}, {Macci{\`o}}, {Somerville}, {Naab}
  \& {Cox}}{{Moster} et~al.}{2011}]{2011MNRAS.415.3750M}
{Moster} B.~P.,  {Macci{\`o}} A.~V.,  {Somerville} R.~S.,  {Naab} T.,   {Cox}
  T.~J.,  2011, \mn@doi [\mnras] {10.1111/j.1365-2966.2011.18984.x}, \href
  {https://ui.adsabs.harvard.edu/abs/2011MNRAS.415.3750M} {415, 3750}

\bibitem[\protect\citeauthoryear{{Moster}, {Naab}  \& {White}}{{Moster}
  et~al.}{2013}]{2013MNRAS.428.3121M}
{Moster} B.~P.,  {Naab} T.,   {White} S. D.~M.,  2013, \mn@doi [\mnras]
  {10.1093/mnras/sts261}, \href
  {https://ui.adsabs.harvard.edu/abs/2013MNRAS.428.3121M} {428, 3121}

\bibitem[\protect\citeauthoryear{{Mukherjee}, {Bicknell}, {Sutherland}  \&
  {Wagner}}{{Mukherjee} et~al.}{2016}]{2016MNRAS.461..967M}
{Mukherjee} D.,  {Bicknell} G.~V.,  {Sutherland} R.,   {Wagner} A.,  2016,
  \mn@doi [\mnras] {10.1093/mnras/stw1368}, \href
  {https://ui.adsabs.harvard.edu/abs/2016MNRAS.461..967M} {461, 967}

\bibitem[\protect\citeauthoryear{{Nesvadba} et~al.,}{{Nesvadba}
  et~al.}{2010}]{2010A&A...521A..65N}
{Nesvadba} N.~P.~H.,  et~al., 2010, \mn@doi [\aap]
  {10.1051/0004-6361/200913333}, \href
  {https://ui.adsabs.harvard.edu/abs/2010A&A...521A..65N} {521, A65}

\bibitem[\protect\citeauthoryear{{Nesvadba} et~al.,}{{Nesvadba}
  et~al.}{2021}]{2021A&A...654A...8N}
{Nesvadba} N.~P.~H.,  et~al., 2021, \mn@doi [\aap]
  {10.1051/0004-6361/202140544}, \href
  {https://ui.adsabs.harvard.edu/abs/2021A&A...654A...8N} {654, A8}

\bibitem[\protect\citeauthoryear{{Nicastro} et~al.,}{{Nicastro}
  et~al.}{2021}]{2021ExA....51.1013N}
{Nicastro} F.,  et~al., 2021, \mn@doi [Experimental Astronomy]
  {10.1007/s10686-021-09710-2}, \href
  {https://ui.adsabs.harvard.edu/abs/2021ExA....51.1013N} {51, 1013}

\bibitem[\protect\citeauthoryear{{Noll}, {Burgarella}, {Giovannoli}, {Buat},
  {Marcillac}  \& {Mu{\~n}oz-Mateos}}{{Noll}
  et~al.}{2009}]{2009A&A...507.1793N}
{Noll} S.,  {Burgarella} D.,  {Giovannoli} E.,  {Buat} V.,  {Marcillac} D.,
  {Mu{\~n}oz-Mateos} J.~C.,  2009, \mn@doi [\aap]
  {10.1051/0004-6361/200912497}, \href
  {https://ui.adsabs.harvard.edu/abs/2009A&A...507.1793N} {507, 1793}

\bibitem[\protect\citeauthoryear{{Nyland} et~al.,}{{Nyland}
  et~al.}{2018}]{2018ApJ...859...23N}
{Nyland} K.,  et~al., 2018, \mn@doi [\apj] {10.3847/1538-4357/aab3d1}, \href
  {https://ui.adsabs.harvard.edu/abs/2018ApJ...859...23N} {859, 23}

\bibitem[\protect\citeauthoryear{Ogle, Lanz, Appleton, Helou  \&
  Mazzarella}{Ogle et~al.}{2019a}]{Ogle_2019}
Ogle P.~M.,  Lanz L.,  Appleton P.~N.,  Helou G.,   Mazzarella J.,  2019a,
  \mn@doi [The Astrophysical Journal Supplement Series]
  {10.3847/1538-4365/ab21c3}, 243, 14

\bibitem[\protect\citeauthoryear{{Ogle}, {Jarrett}, {Lanz}, {Cluver},
  {Alatalo}, {Appleton}  \& {Mazzarella}}{{Ogle}
  et~al.}{2019b}]{2019ApJ...884L..11O}
{Ogle} P.~M.,  {Jarrett} T.,  {Lanz} L.,  {Cluver} M.,  {Alatalo} K.,
  {Appleton} P.~N.,   {Mazzarella} J.~M.,  2019b, \mn@doi [\apjl]
  {10.3847/2041-8213/ab459e}, \href
  {https://ui.adsabs.harvard.edu/abs/2019ApJ...884L..11O} {884, L11}

\bibitem[\protect\citeauthoryear{{Orellana} et~al.,}{{Orellana}
  et~al.}{2017}]{2017A&A...602A..68O}
{Orellana} G.,  et~al., 2017, \mn@doi [\aap] {10.1051/0004-6361/201629009},
  \href {https://ui.adsabs.harvard.edu/abs/2017A&A...602A..68O} {602, A68}

\bibitem[\protect\citeauthoryear{{Papastergis}, {Adams}  \& {van der
  Hulst}}{{Papastergis} et~al.}{2016}]{2016A&A...593A..39P}
{Papastergis} E.,  {Adams} E.~A.~K.,   {van der Hulst} J.~M.,  2016, \mn@doi
  [\aap] {10.1051/0004-6361/201628410}, \href
  {https://ui.adsabs.harvard.edu/abs/2016A&A...593A..39P} {593, A39}

\bibitem[\protect\citeauthoryear{{Peng}, {Ho}, {Impey}  \& {Rix}}{{Peng}
  et~al.}{2002}]{2002AJ....124..266P}
{Peng} C.~Y.,  {Ho} L.~C.,  {Impey} C.~D.,   {Rix} H.-W.,  2002, \mn@doi [\aj]
  {10.1086/340952}, \href
  {https://ui.adsabs.harvard.edu/abs/2002AJ....124..266P} {124, 266}

\bibitem[\protect\citeauthoryear{{Ponomareva} et~al.,}{{Ponomareva}
  et~al.}{2021}]{2021MNRAS.508.1195P}
{Ponomareva} A.~A.,  et~al., 2021, \mn@doi [\mnras] {10.1093/mnras/stab2654},
  \href {https://ui.adsabs.harvard.edu/abs/2021MNRAS.508.1195P} {508, 1195}

\bibitem[\protect\citeauthoryear{{Posti}, {Fraternali}  \& {Marasco}}{{Posti}
  et~al.}{2019}]{2019A&A...626A..56P}
{Posti} L.,  {Fraternali} F.,   {Marasco} A.,  2019, \mn@doi [\aap]
  {10.1051/0004-6361/201935553}, \href
  {https://ui.adsabs.harvard.edu/abs/2019A&A...626A..56P} {626, A56}

\bibitem[\protect\citeauthoryear{{Reda}, {Forbes}, {Beasley}, {O'Sullivan}  \&
  {Goudfrooij}}{{Reda} et~al.}{2004}]{2004MNRAS.354..851R}
{Reda} F.~M.,  {Forbes} D.~A.,  {Beasley} M.~A.,  {O'Sullivan} E.~J.,
  {Goudfrooij} P.,  2004, \mn@doi [\mnras] {10.1111/j.1365-2966.2004.08250.x},
  \href {https://ui.adsabs.harvard.edu/abs/2004MNRAS.354..851R} {354, 851}

\bibitem[\protect\citeauthoryear{{R{\'e}my-Ruyer} et~al.,}{{R{\'e}my-Ruyer}
  et~al.}{2014}]{2014A&A...563A..31R}
{R{\'e}my-Ruyer} A.,  et~al., 2014, \mn@doi [\aap]
  {10.1051/0004-6361/201322803}, \href
  {https://ui.adsabs.harvard.edu/abs/2014A&A...563A..31R} {563, A31}

\bibitem[\protect\citeauthoryear{{Roberts}}{{Roberts}}{1978}]{1978AJ.....83.1026R}
{Roberts} M.~S.,  1978, \mn@doi [\aj] {10.1086/112287}, \href
  {https://ui.adsabs.harvard.edu/abs/1978AJ.....83.1026R} {83, 1026}

\bibitem[\protect\citeauthoryear{{Rowlands} et~al.,}{{Rowlands}
  et~al.}{2012}]{2012MNRAS.419.2545R}
{Rowlands} K.,  et~al., 2012, \mn@doi [\mnras]
  {10.1111/j.1365-2966.2011.19905.x}, \href
  {https://ui.adsabs.harvard.edu/abs/2012MNRAS.419.2545R} {419, 2545}

\bibitem[\protect\citeauthoryear{{Sabater} et~al.,}{{Sabater}
  et~al.}{2019}]{2019A&A...622A..17S}
{Sabater} J.,  et~al., 2019, \mn@doi [\aap] {10.1051/0004-6361/201833883},
  \href {https://ui.adsabs.harvard.edu/abs/2019A&A...622A..17S} {622, A17}

\bibitem[\protect\citeauthoryear{{Sahu}, {Graham}  \& {Davis}}{{Sahu}
  et~al.}{2019}]{2019ApJ...876..155S}
{Sahu} N.,  {Graham} A.~W.,   {Davis} B.~L.,  2019, \mn@doi [\apj]
  {10.3847/1538-4357/ab0f32}, \href
  {https://ui.adsabs.harvard.edu/abs/2019ApJ...876..155S} {876, 155}

\bibitem[\protect\citeauthoryear{{Salim} et~al.,}{{Salim}
  et~al.}{2016}]{2016ApJS..227....2S}
{Salim} S.,  et~al., 2016, \mn@doi [\apjs] {10.3847/0067-0049/227/1/2}, \href
  {https://ui.adsabs.harvard.edu/abs/2016ApJS..227....2S} {227, 2}

\bibitem[\protect\citeauthoryear{{Sanders}, {Soifer}, {Elias}, {Madore},
  {Matthews}, {Neugebauer}  \& {Scoville}}{{Sanders}
  et~al.}{1988}]{1988ApJ...325...74S}
{Sanders} D.~B.,  {Soifer} B.~T.,  {Elias} J.~H.,  {Madore} B.~F.,  {Matthews}
  K.,  {Neugebauer} G.,   {Scoville} N.~Z.,  1988, \mn@doi [\apj]
  {10.1086/165983}, \href
  {https://ui.adsabs.harvard.edu/abs/1988ApJ...325...74S} {325, 74}

\bibitem[\protect\citeauthoryear{{Schaye} et~al.,}{{Schaye}
  et~al.}{2015}]{2015MNRAS.446..521S}
{Schaye} J.,  et~al., 2015, \mn@doi [\mnras] {10.1093/mnras/stu2058}, \href
  {https://ui.adsabs.harvard.edu/abs/2015MNRAS.446..521S} {446, 521}

\bibitem[\protect\citeauthoryear{{Seibert} et~al.,}{{Seibert}
  et~al.}{2012}]{2012AAS...21934001S}
{Seibert} M.,  et~al., 2012, in American Astronomical Society Meeting Abstracts
  \#219. p. 340.01

\bibitem[\protect\citeauthoryear{{Simionescu} et~al.,}{{Simionescu}
  et~al.}{2021}]{2021ExA....51.1043S}
{Simionescu} A.,  et~al., 2021, \mn@doi [Experimental Astronomy]
  {10.1007/s10686-021-09720-0}, \href
  {https://ui.adsabs.harvard.edu/abs/2021ExA....51.1043S} {51, 1043}

\bibitem[\protect\citeauthoryear{{Sofue} \& {Rubin}}{{Sofue} \&
  {Rubin}}{2001}]{2001ARA&A..39..137S}
{Sofue} Y.,  {Rubin} V.,  2001, \mn@doi [\araa]
  {10.1146/annurev.astro.39.1.137}, \href
  {https://ui.adsabs.harvard.edu/abs/2001ARA&A..39..137S} {39, 137}

\bibitem[\protect\citeauthoryear{{Somerville} \& {Dav{\'e}}}{{Somerville} \&
  {Dav{\'e}}}{2015}]{2015ARA&A..53...51S}
{Somerville} R.~S.,  {Dav{\'e}} R.,  2015, \mn@doi [\araa]
  {10.1146/annurev-astro-082812-140951}, \href
  {https://ui.adsabs.harvard.edu/abs/2015ARA&A..53...51S} {53, 51}

\bibitem[\protect\citeauthoryear{{Somerville}, {Hopkins}, {Cox}, {Robertson}
  \& {Hernquist}}{{Somerville} et~al.}{2008}]{2008MNRAS.391..481S}
{Somerville} R.~S.,  {Hopkins} P.~F.,  {Cox} T.~J.,  {Robertson} B.~E.,
  {Hernquist} L.,  2008, \mn@doi [\mnras] {10.1111/j.1365-2966.2008.13805.x},
  \href {https://ui.adsabs.harvard.edu/abs/2008MNRAS.391..481S} {391, 481}

\bibitem[\protect\citeauthoryear{{Sommer-Larsen}}{{Sommer-Larsen}}{2006}]{2006ApJ...644L...1S}
{Sommer-Larsen} J.,  2006, \mn@doi [\apjl] {10.1086/505489}, \href
  {https://ui.adsabs.harvard.edu/abs/2006ApJ...644L...1S} {644, L1}

\bibitem[\protect\citeauthoryear{{Spergel} et~al.,}{{Spergel}
  et~al.}{2007}]{2007ApJS..170..377S}
{Spergel} D.~N.,  et~al., 2007, \mn@doi [\apjs] {10.1086/513700}, \href
  {https://ui.adsabs.harvard.edu/abs/2007ApJS..170..377S} {170, 377}

\bibitem[\protect\citeauthoryear{{Stassun} et~al.,}{{Stassun}
  et~al.}{2019}]{2019AJ....158..138S}
{Stassun} K.~G.,  et~al., 2019, \mn@doi [\aj] {10.3847/1538-3881/ab3467}, \href
  {https://ui.adsabs.harvard.edu/abs/2019AJ....158..138S} {158, 138}

\bibitem[\protect\citeauthoryear{{Sun}, {Voit}, {Donahue}, {Jones}, {Forman}
  \& {Vikhlinin}}{{Sun} et~al.}{2009}]{2009ApJ...693.1142S}
{Sun} M.,  {Voit} G.~M.,  {Donahue} M.,  {Jones} C.,  {Forman} W.,
  {Vikhlinin} A.,  2009, \mn@doi [\apj] {10.1088/0004-637X/693/2/1142}, \href
  {https://ui.adsabs.harvard.edu/abs/2009ApJ...693.1142S} {693, 1142}

\bibitem[\protect\citeauthoryear{{Tody}}{{Tody}}{1986}]{1986SPIE..627..733T}
{Tody} D.,  1986, in {Crawford} D.~L.,  ed.,  Society of Photo-Optical
  Instrumentation Engineers (SPIE) Conference Series Vol. 627, Instrumentation
  in astronomy VI. p.~733, \mn@doi{10.1117/12.968154}

\bibitem[\protect\citeauthoryear{{Tody}}{{Tody}}{1993}]{1993ASPC...52..173T}
{Tody} D.,  1993, in {Hanisch} R.~J.,  {Brissenden} R.~J.~V.,   {Barnes} J.,
  eds,  Astronomical Society of the Pacific Conference Series Vol. 52,
  Astronomical Data Analysis Software and Systems II. p.~173

\bibitem[\protect\citeauthoryear{{Tumlinson}, {Peeples}  \& {Werk}}{{Tumlinson}
  et~al.}{2017}]{2017ARA&A..55..389T}
{Tumlinson} J.,  {Peeples} M.~S.,   {Werk} J.~K.,  2017, \mn@doi [\araa]
  {10.1146/annurev-astro-091916-055240}, \href
  {https://ui.adsabs.harvard.edu/abs/2017ARA&A..55..389T} {55, 389}

\bibitem[\protect\citeauthoryear{{Verheijen}}{{Verheijen}}{2001}]{2001ApJ...563..694V}
{Verheijen} M. A.~W.,  2001, \mn@doi [\apj] {10.1086/323887}, \href
  {https://ui.adsabs.harvard.edu/abs/2001ApJ...563..694V} {563, 694}

\bibitem[\protect\citeauthoryear{{Vikhlinin}, {Kravtsov}, {Forman}, {Jones},
  {Markevitch}, {Murray}  \& {Van Speybroeck}}{{Vikhlinin}
  et~al.}{2006}]{2006ApJ...640..691V}
{Vikhlinin} A.,  {Kravtsov} A.,  {Forman} W.,  {Jones} C.,  {Markevitch} M.,
  {Murray} S.~S.,   {Van Speybroeck} L.,  2006, \mn@doi [\apj]
  {10.1086/500288}, \href
  {https://ui.adsabs.harvard.edu/abs/2006ApJ...640..691V} {640, 691}

\bibitem[\protect\citeauthoryear{{Vogelsberger}, {Marinacci}, {Torrey}  \&
  {Puchwein}}{{Vogelsberger} et~al.}{2020}]{2020NatRP...2...42V}
{Vogelsberger} M.,  {Marinacci} F.,  {Torrey} P.,   {Puchwein} E.,  2020,
  \mn@doi [Nature Reviews Physics] {10.1038/s42254-019-0127-2}, \href
  {https://ui.adsabs.harvard.edu/abs/2020NatRP...2...42V} {2, 42}

\bibitem[\protect\citeauthoryear{{Walker}, {Bagchi}  \& {Fabian}}{{Walker}
  et~al.}{2015}]{2015MNRAS.449.3527W}
{Walker} S.~A.,  {Bagchi} J.,   {Fabian} A.~C.,  2015, \mn@doi [\mnras]
  {10.1093/mnras/stv616}, \href
  {https://ui.adsabs.harvard.edu/abs/2015MNRAS.449.3527W} {449, 3527}

\bibitem[\protect\citeauthoryear{{Wall}, {Kilic}, {Bergeron}, {Rolland},
  {Genest-Beaulieu}  \& {Gianninas}}{{Wall} et~al.}{2019}]{Walletal2019}
{Wall} R.~E.,  {Kilic} M.,  {Bergeron} P.,  {Rolland} B.,  {Genest-Beaulieu}
  C.,   {Gianninas} A.,  2019, \mn@doi [\mnras] {10.1093/mnras/stz2506}, \href
  {https://ui.adsabs.harvard.edu/abs/2019MNRAS.489.5046W} {489, 5046}

\bibitem[\protect\citeauthoryear{{Wechsler} \& {Tinker}}{{Wechsler} \&
  {Tinker}}{2018}]{2018ARA&A..56..435W}
{Wechsler} R.~H.,  {Tinker} J.~L.,  2018, \mn@doi [\araa]
  {10.1146/annurev-astro-081817-051756}, \href
  {https://ui.adsabs.harvard.edu/abs/2018ARA&A..56..435W} {56, 435}

\bibitem[\protect\citeauthoryear{{White} \& {Frenk}}{{White} \&
  {Frenk}}{1991}]{1991ApJ...379...52W}
{White} S. D.~M.,  {Frenk} C.~S.,  1991, \mn@doi [\apj] {10.1086/170483}, \href
  {https://ui.adsabs.harvard.edu/abs/1991ApJ...379...52W} {379, 52}

\bibitem[\protect\citeauthoryear{{White} \& {Rees}}{{White} \&
  {Rees}}{1978}]{1978MNRAS.183..341W}
{White} S.~D.~M.,  {Rees} M.~J.,  1978, \mn@doi [\mnras]
  {10.1093/mnras/183.3.341}, \href
  {https://ui.adsabs.harvard.edu/abs/1978MNRAS.183..341W} {183, 341}

\bibitem[\protect\citeauthoryear{{Young} \& {Knezek}}{{Young} \&
  {Knezek}}{1989}]{1989ApJ...347L..55Y}
{Young} J.~S.,  {Knezek} P.~M.,  1989, \mn@doi [\apjl] {10.1086/185606}, \href
  {https://ui.adsabs.harvard.edu/abs/1989ApJ...347L..55Y} {347, L55}

\bibitem[\protect\citeauthoryear{{Yuan} \& {Narayan}}{{Yuan} \&
  {Narayan}}{2014}]{2014ARA&A..52..529Y}
{Yuan} F.,  {Narayan} R.,  2014, \mn@doi [\araa]
  {10.1146/annurev-astro-082812-141003}, \href
  {https://ui.adsabs.harvard.edu/abs/2014ARA&A..52..529Y} {52, 529}

\bibitem[\protect\citeauthoryear{{Zaritsky} et~al.,}{{Zaritsky}
  et~al.}{2014}]{2014AJ....147..134Z}
{Zaritsky} D.,  et~al., 2014, \mn@doi [\aj] {10.1088/0004-6256/147/6/134},
  \href {https://ui.adsabs.harvard.edu/abs/2014AJ....147..134Z} {147, 134}

\bibitem[\protect\citeauthoryear{{Zhang}, {Wang}, {Luo}, {Zhang}, {Mo}, {Jing},
  {Yang}  \& {Li}}{{Zhang} et~al.}{2021}]{2021arXiv211204777Z}
{Zhang} Z.,  {Wang} H.,  {Luo} W.,  {Zhang} J.,  {Mo} H.~J.,  {Jing} Y.,
  {Yang} X.,   {Li} H.,  2021, arXiv e-prints, \href
  {https://ui.adsabs.harvard.edu/abs/2021arXiv211204777Z} {p. arXiv:2112.04777}

\bibitem[\protect\citeauthoryear{{da Cunha}, {Charlot}  \& {Elbaz}}{{da Cunha}
  et~al.}{2008}]{2008MNRAS.388.1595D}
{da Cunha} E.,  {Charlot} S.,   {Elbaz} D.,  2008, \mn@doi [\mnras]
  {10.1111/j.1365-2966.2008.13535.x}, \href
  {https://ui.adsabs.harvard.edu/abs/2008MNRAS.388.1595D} {388, 1595}

\bibitem[\protect\citeauthoryear{{da Cunha}, {Eminian}, {Charlot}  \&
  {Blaizot}}{{da Cunha} et~al.}{2010}]{2010MNRAS.403.1894D}
{da Cunha} E.,  {Eminian} C.,  {Charlot} S.,   {Blaizot} J.,  2010, \mn@doi
  [\mnras] {10.1111/j.1365-2966.2010.16344.x}, \href
  {https://ui.adsabs.harvard.edu/abs/2010MNRAS.403.1894D} {403, 1894}

\bibitem[\protect\citeauthoryear{{van den Bosch}, {Gebhardt}, {G{\"u}ltekin},
  {Y{\i}ld{\i}r{\i}m}  \& {Walsh}}{{van den Bosch}
  et~al.}{2015}]{2015ApJS..218...10V}
{van den Bosch} R. C.~E.,  {Gebhardt} K.,  {G{\"u}ltekin} K.,
  {Y{\i}ld{\i}r{\i}m} A.,   {Walsh} J.~L.,  2015, \mn@doi [\apjs]
  {10.1088/0067-0049/218/1/10}, \href
  {https://ui.adsabs.harvard.edu/abs/2015ApJS..218...10V} {218, 10}

\makeatother
\end{thebibliography}




\appendix




\bsp	
\label{lastpage}
\end{document}